\documentclass[%
 reprint,
superscriptaddress,
 amsmath,amssymb,
 aps,
 prx,
 longbibliography,
]{revtex4-2}
\usepackage{CJK}
\usepackage{graphicx}
\usepackage{dcolumn}
\usepackage{amssymb}
\usepackage{amsbsy}
\usepackage{amsmath}
\usepackage{mathrsfs}
\usepackage{epsfig}
\usepackage{graphicx}
\usepackage{array}
\usepackage{textcomp}
\usepackage{color}
\usepackage{braket}
\usepackage{bm}
\usepackage[justification=raggedright,font=small]{caption}
\usepackage[titletoc,toc,title]{appendix}
\usepackage[normalem]{ulem}
\usepackage{tikz}

\usepackage{enumitem}
\usepackage[unicode=true,pdfusetitle,
 bookmarks=true,bookmarksnumbered=false,bookmarksopen=false,
 breaklinks=false,pdfborder={0 0 0},pdfborderstyle={},backref=false,colorlinks=true]
 {hyperref}
 
\hypersetup{
 linkcolor=blue, urlcolor=blue, citecolor=blue} 

\edef\flag{1}

\begin{document}
\title{Sequential Generation of Two-dimensional Super-area-law States with Local Parent Hamiltonian}

\author{Wucheng Zhang}
\email{wz8361@princeton.edu}
\affiliation{Department of Physics, Princeton University, Princeton, New Jersey 08544, USA}

\date{\today}

\begin{abstract}
We construct examples of highly entangled two-dimensional states by exploiting a correspondence between stochastic processes in $d$ dimensions and quantum states in $d+1$ dimensions. The entanglement structure of these states, which we explicitly calculate, can be tuned between area law, sub-volume law, and volume law. This correspondence also enables a sequential generation protocol: the states can be prepared through a series of unitary transformations acting on an auxiliary system. We also discuss the conditions under which these states have local, frustration-free parent Hamiltonians. 
\end{abstract}

\maketitle

\vspace{\baselineskip}

\section{Introduction}
\label{sec:intro}
In many-body physics, quantum states are often analyzed through an information-theoretic lens by examining universal features of how the entanglement entropy scales with the structure of a bipartition. The ground states of one-dimensional gapped local Hamiltonians have area-law entanglement entropy\cite{Hastings_2007}, and the same relation has been partially proved in two dimensions \cite{10.1145/3519935.3519962} and conjectured for higher dimensions \cite{RevModPhys.82.277}. However, the entanglement entropy can deviate by a constant in topologically ordered systems \cite{PhysRevLett.96.110405,PhysRevLett.96.110404}, or even become super-area-law\footnote{This means the entanglement entropy of a subregion $D$ of the state scales faster than the size of $\partial D$.} in gapless states, such as the logarithmic scaling in critical systems \cite{PasqualeCalabrese_2004,PhysRevLett.96.010404,PhysRevLett.96.100503}.

A series of recent works has constructed states with entropy scaling faster than logarithmic growth, by exploiting combinatorial structures in the wavefunction \cite{bravyi2012criticality,movassagh2016supercritical,zhang2017novel,Alexander2021,salberger2018fredkin,Salberger2017}. A notable example of this approach are the Motzkin spin chains \cite{bravyi2012criticality,movassagh2016supercritical,zhang2017novel}, which exhibit tunable entanglement scaling, ranging from area-law to volume-law. Furthermore, these states correspond to computations implementable by a finite-state computation model with a first-in-last-out stack, called push-down automata\cite{gopalakrishnan2023push}. This result broadens the understanding of the connection between entanglement entropy and automata complexity theory \cite{crosswhite2008finite,florido2024regular} and also points out a path to construct highly entangled states by incorporating the structure of complex automata in the wavefunction.

The preparation of super-area-law states in experimental quantum platforms remains challenging. Traditional adiabatic evolution often requires fine-tuned Hamiltonians or nonlocal interactions and is hindered by the vanishing of the energy gap. These difficulties arise because super-area-law states exhibit long-range correlations that cannot be captured by shallow local circuits. In contrast, quench dynamics \cite{Calabrese_2005} and random unitary circuits \cite{PhysRevX.7.031016} enable fast entanglement growth, but the resulting states are uncontrollable. Alternative methods, including multiscale quantum circuits \cite{PRXQuantum.4.030334} and measurement-based feedback protocols \cite{PRXQuantum.3.040337,zhu2023nishimori,foss2023experimental,iqbal2024topological}, are capable of generating states with topological or logarithmic scaling. Recently, \cite{gopalakrishnan2023push} has exploited push-down automata techniques, which enables the preparation of the super-logarithmic-law states through sequential generation \cite{Schon2005,Banuls2008,Wei2022,Osborne2010,Wang2017,Astrakhantsev2022,Barratt2021,FossFeig2021,FossFeig2022,Anand2022}--that is, using an auxiliary system that interacts with each qubit in a specific order, or equivalently as a series of quantum channels. This approach is closely related to the experiments of photonic quantum computing \cite{PhysRevLett.103.113602, PhysRevLett.95.010501, PhysRevLett.105.093601, doi:10.1073/pnas.1711003114}.

However, these previous studies have primarily focused on one-dimensional systems. While higher dimensions offer richer entanglement resources and more intricate structures, there are only a few examples of super-area-law states discovered in two dimensions, which involves critical fermion \cite{PhysRevLett.96.010404,PhysRevLett.96.100503,RevModPhys.82.277} or states that couple multiple Motzkin walks together \cite{balasubramanian20232d,zhang2023coupled,Zhang2022quantum}. Automata theory becomes computationally intractable beyond one dimension, a difficulty already evident in two‑dimensional systems where calculating or preparing area‑law entangled states--such as projected entangled‑pair states (PEPS)--is fundamentally  hard \cite{PhysRevLett.125.210504}. Moreover, standard variational ansatzes like PEPS also fail to efficiently represent super‑area‑law entanglement. A general framework for characterizing these states is still lacking, and practically preparing them through local operations remains an open challenge.

In this work, we present a novel framework that establishes a correspondence between stochastic surface growth models in $d$ dimensions and quantum states in $d+1$ dimensions\footnote{The dimensionality refers to the spatial dimension of the surface and the states.}. This approach allows us to construct super-area-law two-dimensional quantum states with controllable entanglement properties. Specifically, we consider the deposition-evaporation model of surface growth, which provides a parametrized interpolation between Kardar-Parisi-Zhang (KPZ) \cite{kardar1986dynamic} and Edwards-Wilkinson (EW) \cite{edwards1982surface} dynamics. By encoding the classical evolution of this model into a quantum state, we characterize the entanglement structure explicitly, revealing distinct scaling behaviors that correspond to different underlying stochastic processes. Our analysis demonstrates that the entanglement entropy exhibits transitions between area-law, sub-volume-law\footnote{This means the entropy of subregion $D$ scales faster than the size of $\partial D$ but slower than the size of $D$.}, and full volume-law scaling as the model parameters are varied.

One of the key advantages of our approach is that it enables the sequential generation of these states using local quantum channels. This method not only provides a practical scheme for preparing highly entangled states on a quantum processor but also offers insight into the dynamical processes that give rise to volume-law entanglement in natural physical systems.

Beyond state generation, we also explore the conditions under which these entangled states serve as ground states of local frustration-free parent Hamiltonians. By carefully designing local constraints that enforce the local probability structure of the deposition-evaporation dynamics, we construct local Hamiltonians whose ground states coincide with the sequentially generated states. This establishes a direct connection between the stochastic surface growth model and the physics of quantum many-body systems, providing an alternative route to realizing highly entangled quantum phases.

Our findings contribute to the growing understanding of how classical stochastic dynamics can be systematically mapped onto entangled quantum states. This work has potential implications for quantum simulation, tensor network representations of many-body states, and the study of entanglement transitions in higher-dimensional quantum systems. More generally, our results suggest that well-understood statistical mechanics models can serve as valuable tools for constructing and analyzing exotic quantum states with prescribed entanglement properties.

The paper is organized as follows. In Section \ref{sec:2d}, we introduce the two-dimensional deposition-evaporation model and demonstrate how its classical dynamics can be encoded into a quantum state. In Section \ref{sec:scaling}, we analyze the entanglement structure of these states and establish the entanglement phase diagram. Section \ref{sec:H} discusses the construction of local frustration-free parent Hamiltonians, and Section \ref{sec:seq} details our sequential generation protocol. We conclude with a discussion of future directions in Section \ref{sec:conclusion}. We also provide a review of Motzkin states and their sequential generation in Appendix \ref{sec:motzkin1d}.

\section{Two-dimensional Super-area-law States from Deposition-evaporation Model}
\label{sec:2d}
\begin{figure}[t]
    \centering
    \if\flag1\includegraphics[width=0.7\columnwidth]{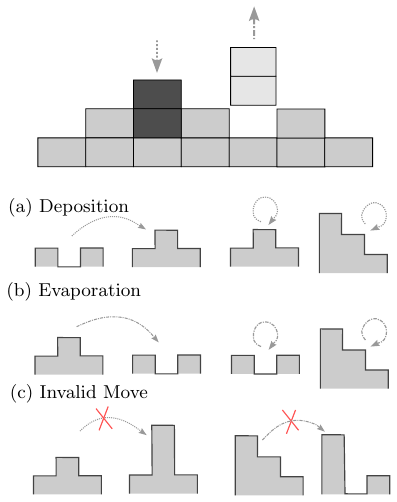}\else\input{demodel}\fi
    \caption{Illustration of the deposition-evaporation model. (a) Under the deposition with probability $p$, while the general intent is to add two blocks, the height difference constraint determines the outcome. Specifically, in the middle and right case where adding two blocks violates the height difference constraint, the only valid deposition leaves the surface invariant, as summarized in \eqref{eq:dep}. (b) Under the evaporation with probability $1-p$, while the general intent is to remove two blocks, the height difference constraint determines the outcome. Specifically, in the middle and right case where removing two blocks violates the height difference constraint, the only valid evaporation leaves the surface invariant, as summarized in \eqref{eq:eva}. (c) The invalid moves that violate the height difference restriction.}
    \label{fig:DEmodel}
\end{figure}

We construct the two-dimensional states by encoding the classical deposition-evaporation model, also known as the entangler-disentangler quantum game \cite{morral2024entanglement},  into the states, as shown in Figure \ref{fig:DEmodel}. This is a surface growth model, extending the idea of the ballistic deposition model \cite{meakin1986ballistic,chame2002crossover,PhysRevE.70.061608,barabasi1995fractal} and introducing ballistic evaporation. The surface is initialized by placing a block at every second site\footnote{More explicitly, $h_i(0)=1$ for $i$ odd, $h_i(0)=0$ for $i$ even, with boundary $h_0(0)=h_{L+1}(0)=0$.}, which we call the \textit{horizon}. Then, at each update step, we randomly select a site and apply either deposition of two blocks with probability $p$ or evaporation of two blocks with probability $1-p$. Let $h_i(t)$ be the height of blocks of site $i \in [0,L+1]$ at time $t$, and $\Delta t$ be the time step of each update. While updating, we also impose the restriction that the height difference of neighboring sites cannot be greater than one:  $|h_i(t)-h_{i+1}(t)|\leq 1$. Therefore, with the initial condition where height differences of the neighboring sites are all one \footnote{In \cite{morral2024entanglement}, the height difference constraint and update rules \eqref{eq:dep} and \eqref{eq:eva} are directly given, with the initial condition as a flat surface.}, the update rule of deposition is given by
\begin{equation}
    h_{i}(t+\Delta t) = \min \{h_{i-1}(t),h_{i+1}(t)\}+1, \label{eq:dep}
\end{equation}
with all three possible cases shown in Figure \ref{fig:DEmodel}(a). And the update rule of evaporation is 
\begin{equation}
    h_{i}(t+\Delta t) = \max \{h_{i-1}(t),h_{i+1}(t)\}-1, \label{eq:eva}
\end{equation}
with all three possible cases shown in Figure \ref{fig:DEmodel}(b). We also impose the restriction $h_i(t) \geq 0$ and the boundary conditions $h_0(t)=h_{L+1}(t)=0$. In the continuum limit $L \to \infty$ and $\Delta t \to 0$, this discrete stochastic process leads to the Kardar–Parisi–Zhang (KPZ) equation \cite{kardar1986dynamic,barabasi1995fractal,morral2024entanglement},
\begin{equation}
    \frac{\partial h}{\partial t}(x,t)=\frac{1}{2}\frac{\partial^2 h}{\partial x^2}-\frac{\lambda_p}{2}\left(\frac{\partial h}{\partial x}\right)^2+\eta(x,t). \label{eq:KPZ}
\end{equation}
where $\eta$ is the Gaussian white noise and the coefficient of the non-linear term $\lambda_p$ has the same sign as $p-1/2$.  \cite{morral2024entanglement} has investigated three phases of this model and extracted the critical exponents. When $p<1/2$, $\lambda_p<0$ stabilizes the roughness and suppresses height growth; the height field of the middle point saturates to $O(1)$. When $p>1/2$, $\lambda_p>0$ gives KPZ dynamics, the middle height $\langle h (L/2,t)\rangle$, where $\langle \cdot\rangle$ means averaging over trajectories of height field configurations, grows as $O(t)$ and saturates to $O(L)$ at time \footnote{This is earlier than the equilibration time of KPZ dynamics due to the open boundary condition and the neighboring height difference constraint.} $T \propto L$. The fluctuation
\begin{equation}
    W(t)=\left\langle\frac{1}{L}\int_0^L (h(x,t)-\bar{h}(t))^2 dx\right\rangle^{1/2},
\end{equation}
where $\bar{h}(t)$ is the averaged height profile over space at time $t$, grows as $O(t^{1/3})$ before the middle height saturates. At the critical point $p_c=1/2$, $\lambda_p=0$ gives Edwards-Wilkinson (EW) \cite{edwards1982surface,barabasi1995fractal,morral2024entanglement} dynamics. The middle height $\langle h(L/2,t)\rangle $ grows as $O(t^{1/4})$ and saturates to $O(\sqrt{L})$ at time $T \propto L^2$. And the fluctuation $W(t)$ grows as $O(t^{1/4})$ and saturates to $O(\sqrt{L})$ at time $T \propto L^2$ as well. 

\begin{figure*}[t]
    \centering
    \if\flag1\includegraphics{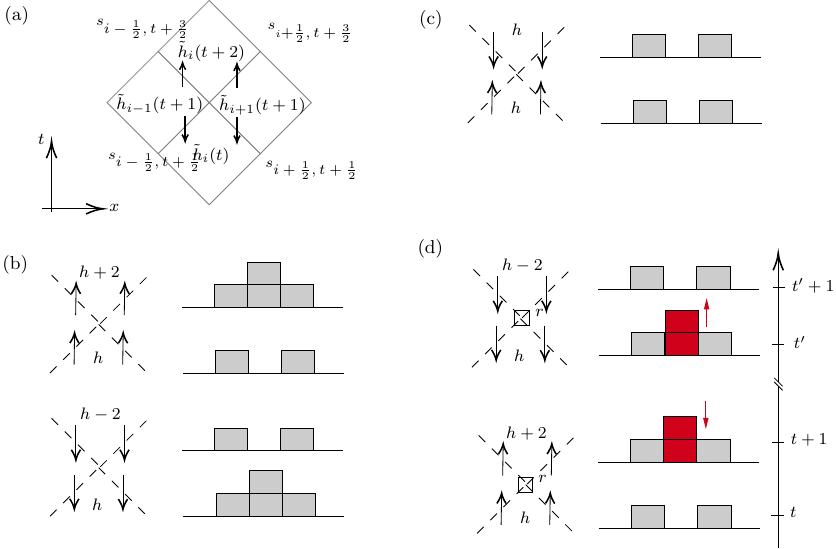}\else\input{stateencode}\fi
    \caption{(a) We place the height fields on the plaquettes and the physical degrees of freedom, spins, on the edges. The horizontal direction labels the spatial direction $i$ of the dynamics, and the vertical direction labels the temporal direction $t$. Spins determine the height field differences. Four spins around a vertex at $(i,t+1)$ define an update from $h_i(t+1)$ to $h_i(t+2)$ while keeping $h_{i\pm1}(t+1)=h_{i\pm1}(t+2)$ unchanged. (b) Examples of spin configurations that correspond to depositing two blocks and evaporating two blocks. (c) Example of a spin configuration with no blocks removed or added. (d) Example of color-matching: The same two red blocks deposited at time $t$ and evaporated at $t'$, then the two vertices corresponding to these two moves should both be labeled as $r$.}
    \label{fig:encoding1}
\end{figure*}
We encode the dynamics into a state on a two-dimensional square lattice with spins placed on edges, first focusing on the uncolored case: 
\begin{equation*}
\vcenter{\hbox{
\tikzset{every picture/.style={line width=0.65pt}} 


}}
\end{equation*}
As we focus primarily on the scaling, we consider a specific shape with $L$ rows and $L$ columns of plaquettes, where $L$ is odd. As demonstrated in Figure \ref{fig:encoding1} (a), the physical $\frac{1}{2}$-spins live on the edges of the lattice and the virtual height field $\Tilde{h}_i(t)$ lives on the plaquettes, where the spatial direction is horizontal and the time direction is vertical. Here we use columns $i=1,\dots,L$ to label the spatial indices of the plaquettes, and rows $t=1,\dots,L$ to label the temporal indices. Naturally, the spins around plaquette $(i,t)$ are labeled by $s_{i\pm\frac{1}{2},t\pm \frac{1}{2}}$. With the same convention, the vertex on column $i$ between plaquettes at $t$ and $t+2$ should be labeled by $(i,t+1)$. The spin gives the height difference of its connecting plaquettes $2s_{i\pm\frac{1}{2},t+\frac{1}{2}}=\tilde{h}_{i\pm 1}(t+1)-\tilde{h}_{i}(t)$. Thus we can impose Gauss's law\footnote{Around each vertex, the sum of the left two spins should equal the sum of the right two spins.} of the four spins around each vertex. As indicated in Figure \ref{fig:encoding1} (b)(c), we use the plaquette height field $\Tilde{h}_i(t)$ in column $i$ row $t$ to encode the surface height profile $h_i(t)$ of site $i$ and time $t$. We can interpret the four spins around vertex $(i,t+1)$ as an update of the blocks at site $i$ happening at time $t+2$ (with no change at $t+1$). Hence, at each time $t$, we update the height at only even or only odd sites, and $h_i(t+1)=h_i(t)=\tilde{h}_i(t)$ if $h_i(t)$ is updated at $t$. Therefore, an equal-time height profile can be read from the zigzag-shaped spatial slice of the lattice. With these rules, we take the superposition of spin configurations encoding the deposition-evaporation dynamics,
\begin{equation}
    |\Psi_\text{uncolored}\rangle = \sum_{\text{trajectory } \alpha}\sqrt{p_\alpha}\,\,\,\vcenter{\hbox{
\tikzset{every picture/.style={line width=0.65pt}} 


    }} \label{eq:configsum}
\end{equation}
where $p_\alpha$ \footnote{{ $p_\alpha$ depends on the entire trajectory $\alpha$, which records the number of sites corresponding to no event, deposition, evaporation, and the special evaporation at $h=0$. As discussed in the rest of the section, under absorbing boundary conditions, where the trajectories are post-selected and reweighted, no explicit closed-form expression exists, while under reflecting boundary conditions $p_\alpha$ is simply the product of the local probabilities specified in the update rules \eqref{eq:UF1}-\eqref{eq:UE6}. For this reason we do not write a single explicit functional form of $p$, the tunable parameter.}} is the probability of the trajectory $\alpha$ in the surface height configuration space $h_i(t)$ and the dots represent the physical spin constrained by the height fields. Here each configuration in \eqref{eq:configsum} is constructed by the local dynamics: at each site, the deposition or evaporation process occurs with a probability of $1/2$, and likewise, the probability of no event occurring is also $1/2$; when a process occurs, the probability of deposition is $p$ and of evaporation is $1-p$. In this process, the configuration should obey boundary conditions $h_0(t)=h_{L+1}(t)=0$, corresponding the spins in the left- and right-most columns of spins ($t=1,3,\dots,L$):
\begin{equation}
    \begin{aligned}
        &s_{\frac{1}{2},t- \frac{1}{2}}= \uparrow,\quad &s_{\frac{1}{2},t+ \frac{1}{2}}= \downarrow,\\
        &s_{L+\frac{1}{2},t-\frac{1}{2}}=\uparrow,\quad &s_{L+\frac{1}{2},t+\frac{1}{2}}=\downarrow 
    \end{aligned}
    \label{eq:leftrightspins}
\end{equation}
We also impose the initial condition at the horizon, and in the end post-select the final surface back to the horizon. The initial and final conditions also constrain the spins in the bottom and top rows ($i=1,3,\dots,L$):
\begin{equation}
    s_{i\pm \frac{1}{2},\frac{1}{2}}=\uparrow,\quad  s_{i\pm \frac{1}{2},L+\frac{1}{2}}=\downarrow.\label{eq:initialfinalspin}
\end{equation}
We require height $h_i(t)\geq 0$, which will be discussed in the end of this section. Finally, we associate the weight of the configuration with its probability $p_\alpha$ generated by the dynamics of the model and normalized among the post-selected trajectories of the height field configuration. 

Optionally, we can add the color-matching condition to obtain $|\Psi_{\text{colored}}\rangle$, i.e., encoding the color into the vertices so that the color of a vertex reflects the color of the blocks evaporated or deposited, as indicated by the surrounding four spins. This is illustrated in Figure \ref{fig:encoding1}(d), where the color of the vertex evaporating two blocks should be matched with the color of the vertex depositing the same blocks. To do so, we can assign the color spin-1 qutrit $c$ to the vertex. When the vertex does not involve evaporated or deposited blocks, we assign $c=0$. When the vertex involves evaporation or deposition of blocks, we assign the color $c=r,g$ (red, green) with equal probability, and it should be the same as the color of the blocks it evaporates or deposits. Then we use the superposition of all such spin configurations, 
\begin{equation}
    |\Psi_{\text{colored}}\rangle = \sum_{\text{trajectory } \alpha}\sqrt{p_\alpha}\,\,\,\vcenter{\hbox{
    \tikzset{every picture/.style={line width=0.65pt}} 



    }} \label{eq:configsumcolor}
\end{equation}
where the dots on edges represent the physical spin-$\frac{1}{2}$ and boxes on vertices represent the physical color spin-1.

We emphasize that there are two ways to achieve the height restriction $h_i(t) \geq 0$: (a) reflecting boundary condition, i.e. at $h_i(t)=0,1$ height cannot be decreased (evaporation might still be allowed as long as it does not decrease the height; but when only deposition is allowed, the probability of deposition rescales from $p$ to 1); or (b) absorbing boundary condition, i.e. there is no constraint during the evolution but we post-select the states with all $h_i(t) \geq 0$ in the end. Compared to the reflecting boundary condition, the absorbing boundary condition suppresses the probability of the configuration close to the height boundary $h=0$. We denote the states generated through reflecting boundary condition as $|\Psi_{\text{ref,colored}}\rangle$ and $|\Psi_{\text{ref,uncolored}}\rangle$, and the states generated through absorbing condition as $|\Psi_{\text{abs,colored}}\rangle$ and $|\Psi_{\text{abs,uncolored}}\rangle$.

\section{Analysis of the Entanglement Structure}
\label{sec:scaling}
\begin{figure}
    \centering
    \if\flag1\includegraphics{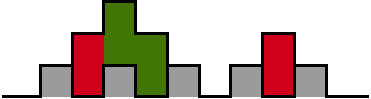}\else\input{entanglement}\fi
    \caption{The visualization of entanglement entropy through the cross-section of a time-slice. The distribution of the contour gives $S_\text{uncolored}$ and the color matching under the contour gives $\langle A\rangle$ area scaling.}
    \label{fig:areaunder}
\end{figure}

In this section, we analyze the scaling behavior of the entanglement entropy and demonstrate the super-area law as we promised. We focus on the colored state $|\Psi_{\text{colored}}\rangle$ and we consider the space-slice cut at $t=L/2$. As we can see in the following analysis, the reflecting and absorbing boundary conditions enforcing $h_i(t) \geq 0$ give the same scaling of entropy, which is mainly determined by the probability distribution of height far away from $h=0$. In the Appendix \ref{app:scalingproof}, we show that with the color matching, any state encoded by classical stochastic dynamics admits a Schmidt decomposition\footnote{The two sets of states $\{|w_L\rangle\}$ and $\{|w_R\rangle\}$ are orthonormal bases because of the fixed initial and final conditions, and color matching .} 
\begin{equation}
    |\Psi_\text{colored}\rangle = \sum_{h_x} \sum_{\Vec{c}_{h_x}} \sqrt{2^{-A_{h_x}}p_{h_x}} |w_{L,h_x}^{\Vec{c}_{h_x}}\rangle \otimes |w_{R,h_x}^{\Vec{c}_{h_x}}\rangle,
\end{equation}
where $p_{h_x}$ is the probability of surface $h_x$ at the cut time $t=L/2$ given by the deposition-evaporation dynamics and boundary conditions, $\Vec{c}_{h_x}$ is the coloring of blocks under the surface $h_x$ (as illustrated in Figure \ref{fig:areaunder}), and $A_{h_x}$ is the area under the surface $h_x$ excluding the blocks from the initial condition. This gives the entanglement entropy \cite{zhang2023coupled}:
\begin{equation}
    S = \alpha\langle A \rangle + S_{\mathrm{uncolored}}, \label{eq:Scontribution}
\end{equation}
where $\langle A \rangle$ is the averaged area under the surface at time $t=L/2$ (as demonstrated in Figure \ref{fig:areaunder}) and $S_{\mathrm{uncolored}}=-\sum_{h_x} p_{h_x} \log p_{h_x}$ is the uncolored contribution. { See more details in the Appendix \ref{app:scalingproof}.} Here we consider the large $L$ limit, which can be described by the dynamics in \eqref{eq:KPZ}.

{To properly evaluate the total entropy $S=\alpha\langle A \rangle + S_{\mathrm{uncolored}}$, we first calculate $S_{\text {uncolored }}$, in order to verify that it does not dominate over the leading term $\alpha\langle A\rangle$. $S_{\mathrm{uncolored}}$ is also the entanglement entropy uncolored state $\left|\Psi_{\text {uncolored}}\right\rangle$.} When $p <1/2$, fluctuation $W(t)$
does not scale with system size $L$ and $h(x)$ is around $O(1)$, thus $S_{\mathrm{uncolored}}$ scales as $O(L)$. At the critical point $p=1/2$, $h(x,t)$ follows Gaussian statistics controlled by fluctuation $W(t) \sim t^{1/4}$ due to the linearity of the dynamics \cite{edwards1982surface,PhysRevE.50.5111}. 
Therefore, the entropy $S_{\mathrm{uncolored}}\sim L \log(W)$ scales as $O(L \log L)$. We do not have to consider the uncolored case with $p > 1/2$. This is because, as shown in the next paragraph, with $p>1/2$ the entropy $S$ of colored states saturates to volume law solely due to $\langle A\rangle$, making it unnecessary to check if $\langle A\rangle$ dominates over $S_\text{uncolored}$.

\begin{figure}
    \centering
    \if\flag1\includegraphics{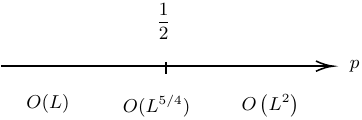}\else\input{phase}\fi
    \caption{The entanglement phase diagram of the state generated by the deposition-evaporation model with color matching.}
    \label{fig:phase}
\end{figure}

Then we consider the contribution from $\langle A\rangle$ to the total bipartite entropy $S$ of the colored states $|\Psi_\text{colored}\rangle$. When $p<1/2$, the surface is confined near $h=0$ and therefore $\langle A\rangle \sim L$ and hence $S \sim L$. At the critical point $p=1/2$, the surface evolves under the EW dynamics and the middle height grows as $\langle h\rangle \propto t^{1/4}$. Then at $t=L/2$, $\langle h\rangle \sim L^{1/4}$ and then $\langle A\rangle \sim L^{5/4}$. Therefore $S \sim L^{5/4}$. Finally, when $p>1/2$, it evolves under the KPZ dynamics, with the middle height growing as $\langle h\rangle \propto t$. And thus at $t = L/2$, $\langle h\rangle \sim L$ and $\langle A\rangle \sim L^2$. Therefore $S \sim L^2$. The phase diagram is summarized in Figure \ref{fig:phase}.

 \section{Construction of the Local Parent Hamiltonian}
\label{sec:H}
To better understand the local structure of the state, we construct a Hamiltonian whose ground state is given by $|\Psi_{\text{abs,colored}}\rangle$\footnote{$|\Psi_{\text{ref}}\rangle$ does not have a local parent Hamiltonian because the Hamiltonian has to detect the superposition of configurations with the surface height touching zero with a different update rule. The height is a non-local information, involving an extensive number of spins along a column, and can only be energetically constrained by nonlocal terms in the Hamiltonian.}, defined in \eqref{eq:configsumcolor}. The uncolored version $|\Psi_{\text{abs,uncolored}}\rangle$ can be obtained by removing the color-matching terms. The construction is based on punishing the superposition that diverges from the probability given by the update rules and boundary conditions. This gives the frustration-free local parent Hamiltonian of $|\Psi_{\text{abs,colored}}\rangle$ in \eqref{eq:configsumcolor}, 
\begin{equation}
   \begin{aligned}
        H = & \sum_{i,t} \left(\sum_{k=1}^3\sum_{c=r,b}\sum_{\Vec{S}}\Pi_{i,t}^{k,c,{\Vec{S}}}\right) + \Pi_{\mathrm{initial}}+ \Pi_{\mathrm{final}}\\
        & + \Pi_{\mathrm{left}} + \Pi_{\mathrm{right}} + \Pi_\text{color} + H_{\text{Gauss}}. \label{eq:parentH}
   \end{aligned}
\end{equation}
Here $\Pi_\text{initial}$, $\Pi_\text{final}$, $\Pi_\text{left}$ and $\Pi_\text{right}$ enforce the boundary conditions in both space and time, while $H_\text{Gauss}$ enforces the Gauss's law around each vertex. Together, these five terms ensure that only spin configurations corresponding to height field trajectories that satisfy the boundary conditions can survive in the ground state. In addition, $\Pi_{\mathrm{color}}$ penalizes invalid color assignments: a vertex that indicates a surface change may not be colored $0$, and a vertex with no change may not be colored $r$ or $g$. Finally, $\Pi_{i,t}^{k,c,\Vec{S}}$ enforces the update rules of deposition-evaporation dynamics and color matching, which will be described in the next paragraph. $\Pi_\text{initial}$ and $\Pi_\text{final}$ require that the surface starts rising from and ends shrinking to the \textit{horizon} configuration, which annihilates the state with \eqref{eq:initialfinalspin}. And $\Pi_\text{left}$ and $\Pi_\text{right}$ require the left- and right-most sites remain with no block deposited, annihilating the state with \eqref{eq:leftrightspins}. The constraints related to initial and final conditions at the \textit{horizon} are also discussed in the one-dimensional examples\cite{bravyi2012criticality,movassagh2016supercritical,zhang2017novel}. The explicit form of these four projectors is given in Appendix \ref{sec:expparentH}. Due to the locality of conditions \eqref{eq:leftrightspins} and \eqref{eq:initialfinalspin}, these four projectors act on the state locally. Similarly, $H_\text{Gauss}$ imposes an energy penalty on any vertex that violates Gauss’s law, while $\Pi_{\text{color}}$ projects onto vertex states exhibiting color violations. They are also local operators, and their explicit forms are provided in Appendix \ref{sec:expparentH}.

For finding the local projector $\Pi_{i,t}^{k,c,\Vec{S}}$, the enforcement of update rules can be constructed from a local deformation from one trajectory to the other, while preserving the surrounding height configuration and color matching. The superposition of these two trajectories is constrained to the local update rules. The minimal deformation is the change of the height $h_i(t)$ on one plaquette, and such a change involves four spins around the plaquette and should also respect another eight spins under Gauss's law. Consider the following illustration as an example,
\begin{equation*}
    \vcenter{\hbox{

\tikzset{every picture/.style={line width=0.75pt}} 

\begin{tikzpicture}[x=0.75pt,y=0.75pt,yscale=-0.8,xscale=0.8]

\draw  [color={rgb, 255:red, 155; green, 155; blue, 155 }  ,draw opacity=1 ][dash pattern={on 4.5pt off 4.5pt}] (104.43,76.97) -- (135.67,108.2) -- (104.43,139.43) -- (73.2,108.2) -- cycle ;
\draw [color={rgb, 255:red, 155; green, 155; blue, 155 }  ,draw opacity=1 ] [dash pattern={on 4.5pt off 4.5pt}]  (104.43,139.43) -- (135.67,170.67) ;
\draw [color={rgb, 255:red, 155; green, 155; blue, 155 }  ,draw opacity=1 ] [dash pattern={on 4.5pt off 4.5pt}]  (73.2,170.67) -- (104.43,139.43) ;
\draw [color={rgb, 255:red, 155; green, 155; blue, 155 }  ,draw opacity=1 ] [dash pattern={on 4.5pt off 4.5pt}]  (73.2,45.73) -- (104.43,76.97) ;
\draw [color={rgb, 255:red, 155; green, 155; blue, 155 }  ,draw opacity=1 ] [dash pattern={on 4.5pt off 4.5pt}]  (104.43,76.97) -- (135.67,45.73) ;
\draw [color={rgb, 255:red, 155; green, 155; blue, 155 }  ,draw opacity=1 ]   (85,165.33) -- (85.3,147) ;
\draw [shift={(85.33,145)}, rotate = 90.94] [color={rgb, 255:red, 155; green, 155; blue, 155 }  ,draw opacity=1 ][line width=0.75]    (10.93,-3.29) .. controls (6.95,-1.4) and (3.31,-0.3) .. (0,0) .. controls (3.31,0.3) and (6.95,1.4) .. (10.93,3.29)   ;
\draw [color={rgb, 255:red, 155; green, 155; blue, 155 }  ,draw opacity=1 ]   (119.88,165.22) -- (120.18,146.88) ;
\draw [shift={(120.22,144.88)}, rotate = 90.94] [color={rgb, 255:red, 155; green, 155; blue, 155 }  ,draw opacity=1 ][line width=0.75]    (10.93,-3.29) .. controls (6.95,-1.4) and (3.31,-0.3) .. (0,0) .. controls (3.31,0.3) and (6.95,1.4) .. (10.93,3.29)   ;
\draw [color={rgb, 255:red, 155; green, 155; blue, 155 }  ,draw opacity=1 ] [dash pattern={on 4.5pt off 4.5pt}]  (41.3,139.77) -- (72.53,108.53) ;
\draw [color={rgb, 255:red, 155; green, 155; blue, 155 }  ,draw opacity=1 ] [dash pattern={on 4.5pt off 4.5pt}]  (135.4,108.53) -- (166.63,77.3) ;
\draw [color={rgb, 255:red, 155; green, 155; blue, 155 }  ,draw opacity=1 ] [dash pattern={on 4.5pt off 4.5pt}]  (135.4,108.53) -- (165.73,137.7) ;
\draw [color={rgb, 255:red, 155; green, 155; blue, 155 }  ,draw opacity=1 ] [dash pattern={on 4.5pt off 4.5pt}]  (42.73,78.33) -- (72.93,108.53) ;
\draw  [fill={rgb, 255:red, 0; green, 0; blue, 0 }  ,fill opacity=1 ] (117.47,92.57) .. controls (117.47,90.82) and (118.88,89.4) .. (120.63,89.4) .. controls (122.38,89.4) and (123.8,90.82) .. (123.8,92.57) .. controls (123.8,94.32) and (122.38,95.73) .. (120.63,95.73) .. controls (118.88,95.73) and (117.47,94.32) .. (117.47,92.57) -- cycle ;
\draw  [fill={rgb, 255:red, 0; green, 0; blue, 0 }  ,fill opacity=1 ] (118.27,122.57) .. controls (118.27,120.82) and (119.68,119.4) .. (121.43,119.4) .. controls (123.18,119.4) and (124.6,120.82) .. (124.6,122.57) .. controls (124.6,124.32) and (123.18,125.73) .. (121.43,125.73) .. controls (119.68,125.73) and (118.27,124.32) .. (118.27,122.57) -- cycle ;
\draw  [fill={rgb, 255:red, 0; green, 0; blue, 0 }  ,fill opacity=1 ] (85.05,92.52) .. controls (85.05,90.77) and (86.47,89.35) .. (88.22,89.35) .. controls (89.97,89.35) and (91.38,90.77) .. (91.38,92.52) .. controls (91.38,94.27) and (89.97,95.68) .. (88.22,95.68) .. controls (86.47,95.68) and (85.05,94.27) .. (85.05,92.52) -- cycle ;
\draw  [fill={rgb, 255:red, 0; green, 0; blue, 0 }  ,fill opacity=1 ] (83.4,121.52) .. controls (83.4,119.77) and (84.82,118.35) .. (86.57,118.35) .. controls (88.32,118.35) and (89.73,119.77) .. (89.73,121.52) .. controls (89.73,123.26) and (88.32,124.68) .. (86.57,124.68) .. controls (84.82,124.68) and (83.4,123.26) .. (83.4,121.52) -- cycle ;
\draw [color={rgb, 255:red, 155; green, 155; blue, 155 }  ,draw opacity=1 ]   (85.4,49.6) -- (85.7,65.93) ;
\draw [shift={(85.73,67.93)}, rotate = 268.96] [color={rgb, 255:red, 155; green, 155; blue, 155 }  ,draw opacity=1 ][line width=0.75]    (10.93,-3.29) .. controls (6.95,-1.4) and (3.31,-0.3) .. (0,0) .. controls (3.31,0.3) and (6.95,1.4) .. (10.93,3.29)   ;
\draw [color={rgb, 255:red, 155; green, 155; blue, 155 }  ,draw opacity=1 ]   (55.8,83.2) -- (56.1,99.53) ;
\draw [shift={(56.13,101.53)}, rotate = 268.96] [color={rgb, 255:red, 155; green, 155; blue, 155 }  ,draw opacity=1 ][line width=0.75]    (10.93,-3.29) .. controls (6.95,-1.4) and (3.31,-0.3) .. (0,0) .. controls (3.31,0.3) and (6.95,1.4) .. (10.93,3.29)   ;
\draw [color={rgb, 255:red, 155; green, 155; blue, 155 }  ,draw opacity=1 ]   (151,115.6) -- (151.3,131.93) ;
\draw [shift={(151.33,133.93)}, rotate = 268.96] [color={rgb, 255:red, 155; green, 155; blue, 155 }  ,draw opacity=1 ][line width=0.75]    (10.93,-3.29) .. controls (6.95,-1.4) and (3.31,-0.3) .. (0,0) .. controls (3.31,0.3) and (6.95,1.4) .. (10.93,3.29)   ;
\draw [color={rgb, 255:red, 155; green, 155; blue, 155 }  ,draw opacity=1 ]   (55.48,131.12) -- (55.78,112.78) ;
\draw [shift={(55.82,110.78)}, rotate = 90.94] [color={rgb, 255:red, 155; green, 155; blue, 155 }  ,draw opacity=1 ][line width=0.75]    (10.93,-3.29) .. controls (6.95,-1.4) and (3.31,-0.3) .. (0,0) .. controls (3.31,0.3) and (6.95,1.4) .. (10.93,3.29)   ;
\draw [color={rgb, 255:red, 155; green, 155; blue, 155 }  ,draw opacity=1 ]   (149.48,99.52) -- (149.78,81.18) ;
\draw [shift={(149.82,79.18)}, rotate = 90.94] [color={rgb, 255:red, 155; green, 155; blue, 155 }  ,draw opacity=1 ][line width=0.75]    (10.93,-3.29) .. controls (6.95,-1.4) and (3.31,-0.3) .. (0,0) .. controls (3.31,0.3) and (6.95,1.4) .. (10.93,3.29)   ;
\draw [color={rgb, 255:red, 74; green, 144; blue, 226 }  ,draw opacity=0.1 ][line width=1.5]    (67.4,54.6) -- (104.2,91.4) -- (136.57,59.03) ;
\draw [shift={(139.4,56.2)}, rotate = 135] [fill={rgb, 255:red, 74; green, 144; blue, 226 }  ,fill opacity=0.1 ][line width=0.08]  [draw opacity=0] (11.07,-5.32) -- (0,0) -- (11.07,5.32) -- (7.35,0) -- cycle    ;
\draw [color={rgb, 255:red, 208; green, 2; blue, 27 }  ,draw opacity=0.1 ][line width=1.5]    (46.6,149) -- (86.6,109) -- (52.23,74.63) ;
\draw [shift={(49.4,71.8)}, rotate = 45] [fill={rgb, 255:red, 208; green, 2; blue, 27 }  ,fill opacity=0.1 ][line width=0.08]  [draw opacity=0] (11.07,-5.32) -- (0,0) -- (11.07,5.32) -- (7.35,0) -- cycle    ;
\draw [color={rgb, 255:red, 155; green, 155; blue, 155 }  ,draw opacity=1 ]   (121,50.8) -- (121.3,67.13) ;
\draw [shift={(121.33,69.13)}, rotate = 268.96] [color={rgb, 255:red, 155; green, 155; blue, 155 }  ,draw opacity=1 ][line width=0.75]    (10.93,-3.29) .. controls (6.95,-1.4) and (3.31,-0.3) .. (0,0) .. controls (3.31,0.3) and (6.95,1.4) .. (10.93,3.29)   ;

\draw (99.2,100) node [anchor=north west][inner sep=0.75pt]    {$h$};

\end{tikzpicture}

    }}
\end{equation*}
where the height of the center plaquette and its surrounding spins (in black dots) are to be deformed, and the other eight gray spins are fixed. Recall that the height difference between plaquettes is always one. Then the center height $h$ can only be deformed to either $h-2$ or $h+2$, depending on the surroundings. To make the deformation possible, three connected plaquettes in the horizontal (e.g., blue arrow) and vertical (e.g., red arrow) directions cannot be in a descending or ascending order in height; otherwise, the deformation would break the height difference constraint. Then the restriction on the middle four spins propagates to the eight surrounding spins through Gauss's law, i.e., the top/bottom two horizontal spins should be the same, and the left/right two vertical spins should be the opposite. Then around plaquette $(i,t)$, once the surrounding spins and involved color $c$ are specified, we can write down the local superposition $|D_{i,t}^{k,c,\Vec{S}}\rangle$, where $k=1,2,3$ labels the cases of allowed deformation by top/bottom four surrounding spins and $\Vec{S}$ label the left/right four surrounding spins. Then the local projector would be in the form
\begin{equation}
   \Pi_{i,t}^{k,c,\Vec{S}} = I_{i,t}^{k,c,\Vec{S}} - \left|D_{i,t}^{k,c,\Vec{S}}\right\rangle\left\langle D_{i,t}^{k,c,\Vec{S}}\right|
\end{equation}
where $I_{i,t}^{k,c,\Vec{S}}$ is the identity only on the subspace of allowed spins and color configurations around plaquette $(i,t)$, respecting the color $c$ matching and Gauss's law constraint from $\Vec{S}$, and the top and bottom spins of case $k$. This local projector annihilates the ground state. It also energetically penalizes states with spin/color configurations around $(i,t)$ that are allowed by $c$ and $\Vec{S}$, but whose superposition violates the prescribed updating probability. The explicit form of $\Pi_{i,t}^{k,c,\Vec{S}}$ and $|D_{i,t}^{k,c,\Vec{S}}\rangle$ are given in Appendix \ref{sec:expparentH}.

\section{Sequential Generation of the State}
\label{sec:seq}
\begin{figure*}[t] 
    \centering
    \begin{tabular}{lc}
         (a)&\\
         &\if\flag1\includegraphics{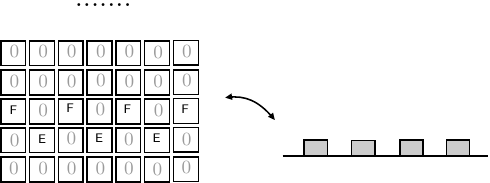}\else\input{stackin}\fi\\
         \\
         (b)&\\
         &\if\flag1\includegraphics{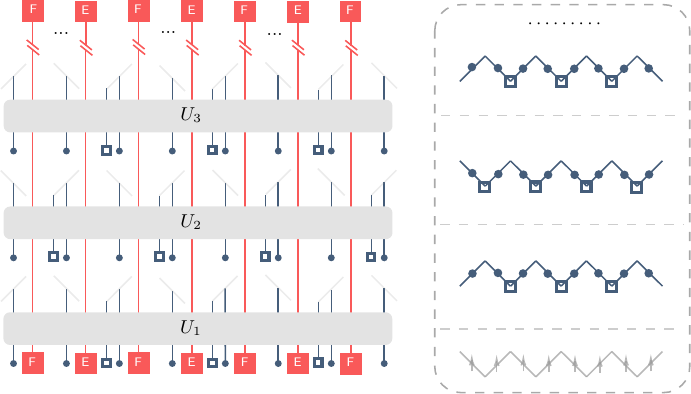}\else\input{stateseq}\fi        
    \end{tabular}
    \caption{(a) The initialized stacks for sequential generation and the corresponding height configuration of the surface. (b) Schematic of sequential generation: Left—sequential application of unitaries $U_n$; right—resulting state aligned with each $U_n$. Spin qubits (blue dots), color qutrits (blue squares), and emitter stacks (red squares labeled $\mathsf{E}/\mathsf{F}$) evolve as follows: red lines trace emitter dynamics, blue lines trace radiated qubits/qutrits. Gray tilted lines denote the lattice geometry of radiated degrees of freedom. At step $n$, $U_n$ acts on the emitter and newly introduced qubits/qutrits, which are then radiated to form a horizontal slice (blue, between dashed lines aligned with $U_n$ and $U_{n+1}$) of the full state. Bottom gray spins represent the fixed initial condition (not part of the generation).}
    \label{fig:stackin}
\end{figure*}
It is hard to approximate $|\Psi_\text{colored}\rangle$ for $p\geq \frac{1}{2}$ cases using projected entangled pair states (PEPS) as the bond dimension grows super-polynomially in $L$. The adiabatic state preparation can involve super-polynomial time complexity, e.g., the inverse energy gap in \cite{zhang2023coupled} is at least $q^{L^3}$ with $q>1$. Motivated by the encoded stochastic process, we can apply the method introduced in \cite{gopalakrishnan2023push} to sequentially generate $|\Psi_{\text{ref,colored}}\rangle$\footnote{We can also extend the procedure to generate $|\Psi_{\text{abs,colored}}\rangle$ by using infinite stacks in both directions. However, the post-selection involves states going below the initial horizon configuration, significantly reducing the efficiency.} in \eqref{eq:configsumcolor}. Recall that $L$ is the number of plaquettes in a row and column. We start with a quantum system with Hilbert space $\mathcal{H}_\text{em}$, called the emitter, which contains $L$ vertical stacks aligned side by side. Each stack is a many-body chain with $O(L)$ sites, where each site has five states labeled by the symbols $0,r,g,\mathsf{E},\mathsf{F}$. In total, we need $O(L^2)$ sites (each with on-site Hilbert space dimension five) for the state with size $L^2$. Each state in $\mathcal{H}_\text{em}$ corresponds to a configuration of deposited blocks, which records the surface growth $h_{i}(t)$ for $1\leq i\leq L$. The emitter is initialized with a reference pure state as in Figure \ref{fig:stackin}(a) corresponding to the initial horizon configuration, where each stack is vertical with $\mathsf{E}$ and $\mathsf{F}$ placed alternatively. In the process of the sequential generation, nontrivial evolution only occurs when $\mathsf{E}(\mathsf{F})$ moves up and down in its own stack, enabling the deposition or evaporation of $r,g$ below $\mathsf{E}(\mathsf{F})$. Figure \ref{fig:stackin}(b) illustrates the sequential application of quantum channels to emit the spin qubits and color qutrits. More precisely, at $n$th application of quantum channels, we bring a reservoir of qubits/qutrits with Hilbert space $\mathcal{H}_n = \mathcal{H}_0^{\otimes (L+1)} \otimes \mathcal{H}_c^{\otimes N_c}$, which is radiated to form the $n$th row of qubits/qutrits of the state. Here $\mathcal{H}_0$ is the Hilbert space of spin-$\frac{1}{2}$ qubit (labeled by $\uparrow$ and $\downarrow$), $\mathcal{H}_c$ is the Hilbert space of color spin-$1$ qutrit (labeled by $0,r$ and $g$), and $N_c$, which denotes the number of color qutrits acted on, is $(L+1)/2$ for even $n$ and $(L+1)/2-1$ for odd $n$. The spin qubits are initialized in state $\downarrow$, and the color qutrits are initialized in state $0$. Then we apply the unitary transformation $U_n$ on $\mathcal{H}_\text{em} \otimes \mathcal{H}_n$. Notice that the bottom row of spins in the lattice is not part of the generation; it merely corresponds to the initial condition of the height evolution and the initial state of the emitter $\mathcal{H}_\text{em}$. In the end, we post-select the outcome in the reference Hilbert space $\mathcal{H}_{\text{em}}$ and separate the desired state in the history $\bigotimes_{n=1}^{L-1} \mathcal{H}_n$\footnote{The top row of the state corresponds to the final condition and is not part of the generation either.}. 

\begin{figure}[t]
    \centering
    \if\flag1\includegraphics{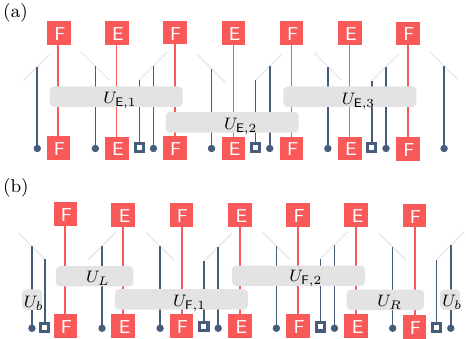}\else\input{fac}\fi
    \caption{(a) Local decomposition of $U_\mathsf{E}$. (b) Local decomposition of $U_{\mathsf{F}}$.}
    \label{fig:Unfactorize}
\end{figure}

When $n$ is odd, $U_n=U_\mathsf{E}$ and we only update stacks with $\mathsf{E}$; when $n$ is even, $U_n=U_\mathsf{F}$ and we only update stacks with $\mathsf{F}$. Locally, each update of $\mathsf{E}/\mathsf{F}$ stacks involves moving $\mathsf{E}/\mathsf{F}$ and rewriting $0$'s and $r,g$'s. To do so, we need to match the configuration of the $\mathsf{E}$($\mathsf{F}$) and its relative position to the other two $\mathsf{F}$($\mathsf{E}$) in the neighboring stacks. As illustrated in Figure \ref{fig:Unfactorize}, $U_{\mathsf{F}(\mathsf{E})}$ acts locally on the three neighboring stacks, as well as two spin qubits and one color qutrit, respectively. We can decompose $U_n$ into local unitary operators, 
\begin{equation}
    U_\mathsf{F}=U_{b}U_{L}U_R\prod_{j=1}^{(L-1)/2-1}U_{\mathsf{F},j}, \quad U_\mathsf{E}=\prod_{j=1}^{(L-1)/2} U_{\mathsf{E},j}.
\end{equation}
With the decomposition, $U_{b}$ only acts on the boundary 1st and $(L+1)$th qubit (and not on the emitter) to enforce the boundary condition, 
\begin{equation}
    \langle \uparrow \uparrow | U_b | \downarrow\downarrow \rangle = 1.
\end{equation}
The operators $U_{L}$ and $U_{R}$ update only the 2nd and $L$th spin qubits, respectively. Each uses data from the boundary stack and its adjacent one, ensuring that the radiated spin qubit encode the height difference between the boundary and its immediate neighbor. The exact forms of $U_{L}$ and $U_{R}$ are given in Appendix \ref{sec:Un}. And $U_{\mathsf{E},j}$ updates only the $2j$th stack (with reference to the $2j-1$th and $2j+1$th stack), the $2j,2j+1$th spin qubits $s_{2j\pm\frac{1}{2},n+\frac{1}{2}}$ as well as the $j$th color qutrit at vertex $(2j,n)$. Finally $U_{\mathsf{F},j}$ updates only the $2j+1$th stack (with reference to the $2j$th and $2j+2$th stack), the $2j,2j+1$th spin qubits $s_{2j+1\pm\frac{1}{2},n+\frac{1}{2}}$ as well as the $j+1$th color qutrit at vertex $(2j+1,n)$. Notice that there are boundary qubits/qutrits already satisfying the boundary condition from initialization, and hence no unitary evolution is needed for them.

\begin{figure}[t]
    \centering
    \if\flag1\includegraphics{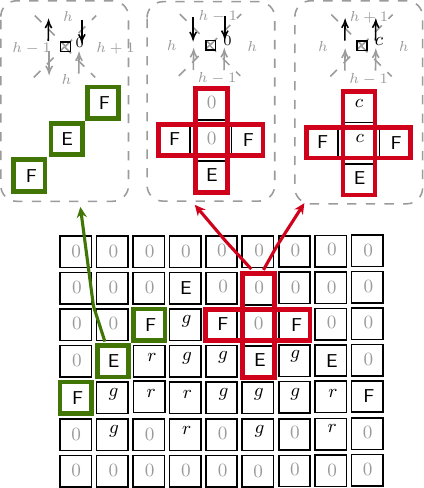}\else\include{stack.tex}\fi
    \caption{Three possible configuration of neighbouring $\mathsf{EFE}$($\mathsf{FEF}$) symbols. The highlighted configuration in the stacks is the state before the application of quantum channels. The arrows point to the dashed boxes of possible moves and radiated reservoir qubits/qutrits (in black, the other symbols in gray are reference information). For demonstration purposes, case (a) is in red and case (c) is in green. For simplicity of the figure, case (b) is not explicitly drawn because it is a reversed version of case (a).}
    \label{fig:statckmove}
\end{figure}

There are three cases of local stacks configuration as shown in three colors in Figure \ref{fig:statckmove}: (a) The three symbols reside in a cross of five plaquettes, where $\mathsf{E}$($\mathsf{F}$) is in the bottom plaquette and two $\mathsf{F}$($\mathsf{E}$) are in the left and right plaquettes; (b) Similar to case (a) but the middle $\mathsf{E}$($\mathsf{F}$) is in the top plaquette; (c) $\mathsf{E}\mathsf{F}\mathsf{E}$($\mathsf{F}\mathsf{E}\mathsf{F}$) symbols are in ascending or descending pattern from left to right. For these three cases, we have different allowed moves: 
\begin{enumerate}[label=(\alph*)]
    \item When $\mathsf{F}$($\mathsf{E}$) is in the bottom plaquette of the cross configuration, we can deposit two blocks and rewrite the middle three plaquettes of the middle stack from $00\mathsf{F}$($00\mathsf{E}$) to $\mathsf{F}cc$($\mathsf{E}cc$) where $c=r,g$; or the stacks can be unchanged and no rewriting is needed. Correspondingly, the radiated qubits/qutrits should be $\uparrow\uparrow$ (for deposition) or $\downarrow\downarrow$ (for no change) and $c$ (for deposition) or 0 (for no change).
    \item When $\mathsf{F}$($\mathsf{E}$) is in the top plaquette of the cross configuration, the stacks can be unchanged with no rewriting; or we can evaporates two blocks and rewrite the middle three plaquettes from $\mathsf{F}cc$($\mathsf{E}cc$) to $00\mathsf{F}$($00\mathsf{E}$) where $c=r,g$. Correspondingly, the radiated qubits/qutrits should be $\downarrow\downarrow$  (for no change) or $\uparrow\uparrow$ (for evaporation) and $0$ (for no change) or $c$ (for evaporation). 
    \item When $\mathsf{E}\mathsf{F}\mathsf{E}$($\mathsf{F}\mathsf{E}\mathsf{F}$) symbols are ascending or descending, no change on the stacks can be made; the corresponding radiated spin qubits are $\uparrow\downarrow$ and $\downarrow\uparrow$ respectively, and the corresponding radiated color qutrit is $0$ for both cases.
\end{enumerate}
According to these, the explicit matrix elements of $U_{\mathsf{E(F)},j}$ are given in Appendix \ref{sec:Un}, where the nonzero elements represent the allowed moves and the exact values are given by the probability of all the events -- no process, deposition, or evaporation -- that contribute to the surface evolution recorded in the stacks change. Notice that we can use the above procedure to generate the uncolored state $|\Psi_\text{ref,uncolored}\rangle$ by restricting $c$ to $0,r$ as only one color.

{ While this sequential scheme ends with a post-selection whose success probability is exponentially small in $L$, we can alternatively make the final check essentially deterministic by adding a brief evaporation-only ``cooling" phase. After the first $L / 2$ rounds, when the space-like cut already encloses a large area, we modify the local update so that deposition is disabled (only ``no event" or evaporation occurs, i.e. $p=0$). This makes the height profile decrease and quickly drains the surface toward the horizon. Running this cooling stage for $L/2$ additional rounds empties the stacks with very large probability, so the final post-selection merely removes rare residual configurations. Operationally, this is implemented by setting the tunable parameter $p=0$ and turning off the deposition branches \eqref{eq:UF3} and \eqref{eq:UE3} in the same local channels; no extra resources are required. Since the entanglement entropy $S\geq \langle A\rangle$ is lower bounded by the surface profile at $t=L/2$, which is unaffected by the cooling phase.}

\section{Conclusion}
\label{sec:conclusion}
In this work, we have explored a novel approach for constructing and analyzing super-area-law quantum states by leveraging the correspondence between classical stochastic processes and quantum many-body wavefunctions. Specifically, we have demonstrated that the classical deposition-evaporation model of surface growth, which interpolates between Kardar–Parisi–Zhang (KPZ) and Edwards–Wilkinson (EW) dynamics, can be encoded into a quantum state, leading to a family of highly entangled two-dimensional states with tunable entanglement scaling. Our results highlight how different stochastic processes govern the entanglement structure, allowing for transitions between area-law, sub-volume-law, and volume-law scaling.

A key feature of our construction is the sequential generation protocol, which enables preparing these entangled states using a unitary circuit model. This approach builds upon the push-down automata framework proposed in \cite{gopalakrishnan2023push} and extends it to two-dimensional systems. Our sequential generation scheme offers a practical method for preparing complex quantum states and suggests potential implementations on near-term quantum devices.

Furthermore, we have established the conditions under which these entangled states can serve as ground states of local frustration-free parent Hamiltonians. By encoding the probability structure of the deposition-evaporation model into local Hamiltonian constraints, we construct explicit Hamiltonians whose ground states coincide with the sequentially generated states. This connection between classical stochastic dynamics and quantum many-body ground states offers a novel perspective for state preparation and Hamiltonian engineering.

In \cite{Zhang2022quantum,zhang2023coupled}, other super-area-law states are constructed with parent Hamiltonians. The states, designed from two-dimensional surface fluctuation and a different color-matching mechanism, have entanglement entropy transiting from area law to volume law with $S \sim L \log L$ at the critical point. The parent Hamiltonians in these two papers are rotational invariant, and hence the entropy scaling is isotropic. Consequently, the Motzkin condition and the color-matching condition are imposed along both the time-slice and the space-slice, which cannot be accommodated by the stack memory, and hence cannot be sequentially generated. In contrast, the states constructed in our work are not isotropic, since the color matching along the space direction is not reinforced, even in the colored case. To see this, we can also consider a time-like cut, and the entanglement at the critical point is mainly contributed by $L$ uncolored and coupled Motzkin chains. Lacking the entanglement contribution \eqref{eq:Scontribution} from the color matching, the entanglement entropy of the time-like cut is lower than that of the space-like cut. Another isotropic construction was presented in \cite{balasubramanian20232d}, where bipartite entanglement at the critical point scales as $S\sim {L^{3/2}}$ and sequential generation is likewise not feasible.

In summary, our findings contribute to the broader understanding of how classical statistical mechanics models can inform the design of quantum states, with implications for quantum computation, entanglement transitions, and tensor network representations of many-body states. { Future directions include extending this framework to a broader range of classical dynamics in various dimensions and investigating potential experimental realizations and quantum computation applications \cite{PhysRevLett.103.113602, PhysRevLett.95.010501, PhysRevLett.105.093601, doi:10.1073/pnas.1711003114}. Notably, in one dimension, deformed Motzkin chains exhibit symmetry-protected topological (SPT) order \cite{PhysRevB.96.180404} and thereby serve as resource states for measurement-based quantum computation (MBQC), where SPT order enables protected teleportation and single-qubit gate operations \cite{PhysRevLett.106.070501,PhysRevLett.108.240505,PhysRevLett.119.010504}, underscoring the potential computational utility of the two-dimensional state family we construct. Our results also suggest a broader principle that entanglement features of the constructed states correspond to classical information quantities of the underlying stochastic trajectories. Exploring such connections, for example, on mutual information, offers a promising direction for future work.}

\subsection*{Acknowledgment}
I am deeply indebted to Sarang Gopalakrishnan for first suggesting the sequential generation approach to highly entangled states and for his many insightful comments that kept this project on the right course. I also thank Hao Chen, J. Alexander Jacoby, Luna Liu, Abhinav Prem, Yifan Zhang, and Zihan Zhou for stimulating discussions and feedback on earlier versions of the manuscript. I am also grateful to the countless cups of free coffee in Jadwin Hall and Bloomberg Hall. This work is supported by the Princeton University Department of Physics. 
\begin{appendix}
\numberwithin{equation}{section}

\section{Motzkin States and their Sequential Generation}
\label{sec:motzkin1d}
In this section, we briefly review the Motzkin state formalism as a warm-up. The Motzkin state \cite{bravyi2012criticality,movassagh2016supercritical,zhang2017novel} is defined within the local Hilbert space $\mathcal{H}_L$ of $L$ spin-$2$ (with states $\pm2,\pm1,0$), formed by the superposition of colored Motzkin walks, i.e.
\begin{equation}
    |\Psi_q \rangle = \frac{1}{\mathcal{N}} \sum_{w \in \text { colored Motzkin walks }} q^{\frac{1}{2} \mathcal{A}(w)}|w\rangle \label{eq:motzkin1d}
\end{equation}
where $\mathcal{N}$ is the normalization constant, $q>0$ a tunable parameter and $\mathcal{A}(w)$ is the area under the Motzkin walks. Motzkin walks are random walks that start and end at the origin and remain above $h=0$, with the up ($+$), down ($-$) spins corresponding to upward and downward moves of one step, and zero (0) spins corresponding to no move, respectively. At a given height, when it increases with the upward spin of the color $c=1,2$, it should decrease with a downward spin of the same color $c$. An example of a Motzkin walk is shown below, where $\pm 1$ spins are in red, $\pm 2$ spins are in green, and $0$ spins are uncolored (black); the colors are matched.
\begin{center}

\tikzset{every picture/.style={line width=0.75pt}} 

\begin{tikzpicture}[x=0.75pt,y=0.75pt,yscale=-1,xscale=1]

\draw [color={rgb, 255:red, 65; green, 117; blue, 5 }  ,draw opacity=1 ]   (79.5,151.5) -- (97.88,133.12) ;
\draw [shift={(100,131)}, rotate = 135] [fill={rgb, 255:red, 65; green, 117; blue, 5 }  ,fill opacity=1 ][line width=0.08]  [draw opacity=0] (6.25,-3) -- (0,0) -- (6.25,3) -- cycle    ;
\draw [color={rgb, 255:red, 208; green, 2; blue, 27 }  ,draw opacity=1 ]   (100,131) -- (118.38,112.62) ;
\draw [shift={(120.5,110.5)}, rotate = 135] [fill={rgb, 255:red, 208; green, 2; blue, 27 }  ,fill opacity=1 ][line width=0.08]  [draw opacity=0] (6.25,-3) -- (0,0) -- (6.25,3) -- cycle    ;
\draw [color={rgb, 255:red, 208; green, 2; blue, 27 }  ,draw opacity=1 ]   (120.5,110.5) -- (138.38,128.38) ;
\draw [shift={(140.5,130.5)}, rotate = 225] [fill={rgb, 255:red, 208; green, 2; blue, 27 }  ,fill opacity=1 ][line width=0.08]  [draw opacity=0] (6.25,-3) -- (0,0) -- (6.25,3) -- cycle    ;
\draw [color={rgb, 255:red, 65; green, 117; blue, 5 }  ,draw opacity=1 ]   (140.5,130.5) -- (158.88,112.12) ;
\draw [shift={(161,110)}, rotate = 135] [fill={rgb, 255:red, 65; green, 117; blue, 5 }  ,fill opacity=1 ][line width=0.08]  [draw opacity=0] (6.25,-3) -- (0,0) -- (6.25,3) -- cycle    ;
\draw [color={rgb, 255:red, 208; green, 2; blue, 27 }  ,draw opacity=1 ]   (161,110) -- (179.38,91.62) ;
\draw [shift={(181.5,89.5)}, rotate = 135] [fill={rgb, 255:red, 208; green, 2; blue, 27 }  ,fill opacity=1 ][line width=0.08]  [draw opacity=0] (6.25,-3) -- (0,0) -- (6.25,3) -- cycle    ;
\draw [color={rgb, 255:red, 208; green, 2; blue, 27 }  ,draw opacity=1 ]   (181.5,89.5) -- (199.38,107.38) ;
\draw [shift={(201.5,109.5)}, rotate = 225] [fill={rgb, 255:red, 208; green, 2; blue, 27 }  ,fill opacity=1 ][line width=0.08]  [draw opacity=0] (6.25,-3) -- (0,0) -- (6.25,3) -- cycle    ;
\draw    (201.5,109.5) -- (224,109.5) ;
\draw [shift={(227,109.5)}, rotate = 180] [fill={rgb, 255:red, 0; green, 0; blue, 0 }  ][line width=0.08]  [draw opacity=0] (6.25,-3) -- (0,0) -- (6.25,3) -- cycle    ;
\draw [color={rgb, 255:red, 65; green, 117; blue, 5 }  ,draw opacity=1 ]   (228,110.5) -- (245.88,128.38) ;
\draw [shift={(248,130.5)}, rotate = 225] [fill={rgb, 255:red, 65; green, 117; blue, 5 }  ,fill opacity=1 ][line width=0.08]  [draw opacity=0] (6.25,-3) -- (0,0) -- (6.25,3) -- cycle    ;
\draw [color={rgb, 255:red, 65; green, 117; blue, 5 }  ,draw opacity=1 ]   (249,131.5) -- (266.88,149.38) ;
\draw [shift={(269,151.5)}, rotate = 225] [fill={rgb, 255:red, 65; green, 117; blue, 5 }  ,fill opacity=1 ][line width=0.08]  [draw opacity=0] (6.25,-3) -- (0,0) -- (6.25,3) -- cycle    ;
\draw [color={rgb, 255:red, 155; green, 155; blue, 155 }  ,draw opacity=1 ] [dash pattern={on 0.84pt off 2.51pt}]  (99,144) -- (250,144) ;
\draw [shift={(252,144)}, rotate = 180] [color={rgb, 255:red, 155; green, 155; blue, 155 }  ,draw opacity=1 ][line width=0.75]    (4.37,-1.32) .. controls (2.78,-0.56) and (1.32,-0.12) .. (0,0) .. controls (1.32,0.12) and (2.78,0.56) .. (4.37,1.32)   ;
\draw [shift={(97,144)}, rotate = 0] [color={rgb, 255:red, 155; green, 155; blue, 155 }  ,draw opacity=1 ][line width=0.75]    (4.37,-1.32) .. controls (2.78,-0.56) and (1.32,-0.12) .. (0,0) .. controls (1.32,0.12) and (2.78,0.56) .. (4.37,1.32)   ;
\draw [color={rgb, 255:red, 155; green, 155; blue, 155 }  ,draw opacity=1 ] [dash pattern={on 0.84pt off 2.51pt}]  (155,125) -- (232,125) ;
\draw [shift={(234,125)}, rotate = 180] [color={rgb, 255:red, 155; green, 155; blue, 155 }  ,draw opacity=1 ][line width=0.75]    (4.37,-1.32) .. controls (2.78,-0.56) and (1.32,-0.12) .. (0,0) .. controls (1.32,0.12) and (2.78,0.56) .. (4.37,1.32)   ;
\draw [shift={(153,125)}, rotate = 0] [color={rgb, 255:red, 155; green, 155; blue, 155 }  ,draw opacity=1 ][line width=0.75]    (4.37,-1.32) .. controls (2.78,-0.56) and (1.32,-0.12) .. (0,0) .. controls (1.32,0.12) and (2.78,0.56) .. (4.37,1.32)   ;
\draw [color={rgb, 255:red, 155; green, 155; blue, 155 }  ,draw opacity=1 ] [dash pattern={on 0.84pt off 2.51pt}]  (112.5,125) -- (127.5,125) ;
\draw [shift={(129.5,125)}, rotate = 180] [color={rgb, 255:red, 155; green, 155; blue, 155 }  ,draw opacity=1 ][line width=0.75]    (4.37,-1.32) .. controls (2.78,-0.56) and (1.32,-0.12) .. (0,0) .. controls (1.32,0.12) and (2.78,0.56) .. (4.37,1.32)   ;
\draw [shift={(110.5,125)}, rotate = 0] [color={rgb, 255:red, 155; green, 155; blue, 155 }  ,draw opacity=1 ][line width=0.75]    (4.37,-1.32) .. controls (2.78,-0.56) and (1.32,-0.12) .. (0,0) .. controls (1.32,0.12) and (2.78,0.56) .. (4.37,1.32)   ;
\draw [color={rgb, 255:red, 155; green, 155; blue, 155 }  ,draw opacity=1 ] [dash pattern={on 0.84pt off 2.51pt}]  (173,105) -- (188,105) ;
\draw [shift={(190,105)}, rotate = 180] [color={rgb, 255:red, 155; green, 155; blue, 155 }  ,draw opacity=1 ][line width=0.75]    (4.37,-1.32) .. controls (2.78,-0.56) and (1.32,-0.12) .. (0,0) .. controls (1.32,0.12) and (2.78,0.56) .. (4.37,1.32)   ;
\draw [shift={(171,105)}, rotate = 0] [color={rgb, 255:red, 155; green, 155; blue, 155 }  ,draw opacity=1 ][line width=0.75]    (4.37,-1.32) .. controls (2.78,-0.56) and (1.32,-0.12) .. (0,0) .. controls (1.32,0.12) and (2.78,0.56) .. (4.37,1.32)   ;

\end{tikzpicture}
\end{center}
We can bipartite the one-dimensional chain at the middle point and calculate the entanglement entropy. In \cite{zhang2017novel} it is shown that when $q>1$, the entanglement entropy follows the volume law $S \sim L$; when $q=1$, the entanglement entropy follows the the sub-volume law $S \sim \sqrt{L}$; and when $q<1$, the entanglement entropy follows the area law. Despite the state’s intricate entanglement structure, local frustration‑free parent Hamiltonians for $|\Psi_q\rangle$ can nevertheless be constructed\cite{bravyi2012criticality,movassagh2016supercritical,zhang2017novel}.

Gopalakrishnan \cite{gopalakrishnan2023push} proposed a method of sequentially generating the state in \eqref{eq:motzkin1d} which has modified weights of trajectories touching $h=0$ but gives the same entanglement scaling, 
\begin{equation}
    |\Psi'\rangle = \sum_{w \in \text { colored Motzkin walks }} \sqrt{p_w} |w\rangle, \label{eq:randomwalk1d}
\end{equation}
where $p_w$ is given by the colored biased random walk in one dimension. Let $p_i(t)$ be the probability of the walk being at height $i$ at time $t$. The transition for $i > 0$ is given by: 
\begin{equation}
    \begin{aligned}
        p_i(t+1)&=\left(1-\gamma_{L}-\gamma_{R}\right) p_i(t)+\gamma_{L} p_{i+1}(t)+\gamma_{R} p_{i-1}(t)\\
    \end{aligned}
\end{equation}
with tunable parameter $\gamma_{L,R}>0$ and $\gamma_L+\gamma_R<1$; and the transition for $i=0$ is controlled by a reflection boundary condition, 
\begin{equation}
    p_i(t+1) = (1-\gamma_0)p_i(t) + \gamma_0 p_{i+1}(t)
\end{equation}
with a tunable parameter $0<\gamma_{0}<1$. The generation method is based on sequentially applying quantum channels on a quantum system with Hilbert space $\mathcal{H}_\mathrm{em}$, called emitter, where each state $|i\rangle$ records the real-time height $i$ and coloring. The emitter is initialized with a reference pure state $|0\rangle \langle 0 |$, representing the initial zero height. The $n$th application of the quantum channel involves applying the unitary transformation $U_n$ on the emitter $\mathcal{H}_{\mathrm{em}}$ and the $n$th reservoir qudit (spin-2) with Hilbert space $\mathcal{H}_n$ that dictates the $n$th move of the walk. After the $n$th application, the $n$th qudit is radiated out and does not interact with the rest of the process. During the whole process, we collect all the reservoir qudits as the physical system. In the end, we post-select the outcome in the reference Hilbert space $\mathcal{H}_{\mathrm{em}}$, where the reference state is desired to return back to the initial zero height of the walk, and then we disentangle reservoir qudits, generating the desired state in the history $\bigotimes_{n=1}^L \mathcal{H}_n$. Here each $U_n$ has no knowledge about system size $L$ and only acts on $\mathcal{H}_{\mathrm{em}}\otimes \mathcal{H}_n$ locally in time. This protocol provides a more efficient generation of super-area law states, as discussed in \cite{gopalakrishnan2023push}. 
\section{KPZ Dynamics from the Deposition-evaporation Model}
We want to derive a stochastic differential equation from the deposition and evaporation process, applying the method developed in \cite{PhysRevE.70.061608}. We define two white Gaussian noise $\eta_i(t)$ and $\chi_i(t)$. We pick $a_p$ such that $\mathbb{E}[\mathbf{1}_{\chi\geq a_p}]=p$. As argued in \cite{PhysRevE.70.061608}, the update rule can be approximated by
\begin{equation}
    h_i(t+1)=\max\{h_{i+1}(t),h_{i-1}(t)\}+\eta_i(t)
\end{equation}
or 
\begin{equation}
    h_i(t+1)=\min\{h_{i+1}(t),h_{i-1}(t)\}+\eta_i(t).
\end{equation}
for each site. Then this process can be written as 
\begin{equation}
    \begin{aligned}
        h_i(t+1) =&\theta(\chi_{i}(t)-a_p)\left(\min\{h_{i+1}(t),h_{i-1}(t)\}+\eta_i(t)\right)\\
        &+\theta(a_p-\chi_{i}(t))\left(\max\{h_{i+1}(t),h_{i-1}(t)\}+\eta_i(t)\right),
    \end{aligned}
\end{equation}
where $\theta(x)$ is the step function. Then we can use the following trick
\begin{equation}
    \begin{aligned}
        \min \{a,b\}&=a\theta(b-a)+b\theta(a-b),\\
        \max \{a,b\}&=a\theta(a-b)+b\theta(b-a),
    \end{aligned}
\end{equation}
and the substitution
\begin{equation}
    \theta(x)=\frac{1}{2}+\frac{1}{2}\mathrm{sgn}(x). 
\end{equation}
And we arrive at the following equation,
\begin{equation}
    \begin{aligned}
        \frac{\partial h}{\partial t}(x, t)=&\frac{(\Delta x)^2}{2 \Delta t}\frac{\partial^2 h}{\partial x^2}(x, t)\\
        & -\mathrm{sgn}(\chi(x,t)-a_p)\frac{\Delta x}{\Delta t}\left|\frac{\partial h}{\partial x}\right|\\
        & +\frac{1}{\Delta t} \eta(x, t)\\
    \end{aligned}
\end{equation}
where $\Delta x$ and $\Delta t$ are finite differences of $x$ and $t$. We can then use the approximation derived in \cite{PhysRevE.70.061608}:
\begin{equation}
    |\nabla h| \simeq \frac{\langle|\nabla h|\rangle}{\left\langle|\nabla h|^2\right\rangle}(\nabla h)^2,
\end{equation}
where $\langle\cdots\rangle$ means steady-state averaging. We can then set $\Delta x=\Delta t=1$ and use $\lambda / 2 \simeq \frac{\langle|\nabla h|\rangle}{\left\langle|\nabla h|^2\right\rangle}$ where $\lambda \geq 0$. Finally, we arrive at the equation,
\begin{equation}
    \frac{\partial h}{\partial t}(x,t)=\frac{1}{2}\frac{\partial^2 h}{\partial x^2}(x,t)-\frac{\lambda}{2}\mathrm{sgn}(\chi(x,t)-a_p)\left(\frac{\partial h}{\partial x}\right)^2+\eta(x,t). \label{eq:kpzlike}
\end{equation}
If $p\to1$, \eqref{eq:kpzlike} reduces to KPZ,
\begin{equation}
    \frac{\partial h}{\partial t}(x,t)=\frac{1}{2}\frac{\partial^2 h}{\partial x^2}(x,t)-\frac{\lambda}{2}\left(\frac{\partial h}{\partial x}\right)^2+\eta(x,t). 
\end{equation}
If $\frac{1}{2}<p<1$, \eqref{eq:kpzlike} is effectively a KPZ
\begin{equation}
     \frac{\partial h}{\partial t}(x,t)=\frac{1}{2}\frac{\partial^2 h}{\partial x^2}(x,t)-\frac{\lambda_p}{2}\left(\frac{\partial h}{\partial x}\right)^2+\eta(x,t).
\end{equation}
where $\lambda_p>0$. If $p=\frac{1}{2}$, \eqref{eq:kpzlike} is effectively EW,
\begin{equation}
     \frac{\partial h}{\partial t}(x,t)=\frac{1}{2}\frac{\partial^2 h}{\partial x^2}(x,t)+\eta(x,t).
\end{equation}
If $0<p<\frac{1}{2}$ is effectively a KPZ
\begin{equation}
     \frac{\partial h}{\partial t}(x,t)=\frac{1}{2}\frac{\partial^2 h}{\partial x^2}(x,t)-\frac{\lambda_p}{2}\left(\frac{\partial h}{\partial x}\right)^2+\eta(x,t).
\end{equation}
where $\lambda_p<0$. In such cases, roughness is stabilized and height growth is suppressed. 
\section{Scaling of Entanglement Entropy with Color Matching}
\label{app:scalingproof}
In this section, we derived the entanglement entropy with color matching in the stochastic process. We assume that at each step, the allowed change in height (increase or decrease) has the same size $|\Delta h|$. We further require that the colors assigned to the change should be matched at the same height. We first consider the one-dimensional case of state-encoded classical dynamics,
\begin{equation}
    |\Psi\rangle= \sum_{\text{uncolored path }i} \sum_{{\Vec{c} \in C_i}} \sqrt{p_i p_{\Vec{c}|i}} |w_i^{\Vec{c}}\rangle, 
\end{equation}
where $w_i^{\Vec{c}}$ is the colored classical path $i$, $p_i$ is the probability of occurrence of such a path given by the classical dynamics (independent of color), $\Vec{c} \in C_i$ represents the coloring of each change along the uncolored path $w_i$ and $p_{\Vec{c}|i}$ represents the probability of the coloring of path $w_i$. To evaluate the entanglement entropy, we apply the Schmidt decomposition by constructing partitioned states from the superposition of left and right half-paths. These paths reach a common height $ h $ at the cut position $ L/2 $, and share a consistent unmatched color profile $ \vec{c}_h $ at the cut:
\begin{equation}
\begin{aligned}
    |w_{L,h}^{\vec{c}_h}\rangle &= \sum_{\substack{i_L \in \mathcal{L}_h \\ \vec{c}_L \in C_{i_L,\vec{c}_h}}} \sqrt{p_{i_L|h} \; p_{\vec{c}_L|\vec{c}_h,i_L}}\,|w_{i_L}^{\vec{c}_L}\rangle, \\
    |w_{R,h}^{\vec{c}_h}\rangle &= \sum_{\substack{i_R \in \mathcal{R}_h \\ \vec{c}_R \in C_{i_R,\vec{c}_h}}} \sqrt{p_{i_R|h} \; p_{\vec{c}_R|\vec{c}_h,i_R}}\,|w_{i_R}^{\vec{c}_R}\rangle,
\end{aligned}
\end{equation}
where $ \mathcal{L}_h $ and $ \mathcal{R}_h $ denote the sets of uncolored left and right half-paths reaching height $ h $ at the cut. The sets $C_{i_L(R),\vec{c}_h}$ include all the colorings of left and right half paths with unmatched color profile $\vec{c}_h$. The vectors $ \vec{c}_L $ and $ \vec{c}_R $ represent colorings of the left and right half-paths, respectively, constrained to match the unmatched color profile $ \vec{c}_h $ at the cut. The term $ p_{i_{L(R)}|h} $ is the probability of a given half-path $ i_{L(R)} $ conditioned on reaching height $ h $, and $ p_{\vec{c}_{L(R)}|\vec{c}_h,i_{L(R)}} $ is the probability of assigning a valid coloring given $ i_{L(R)} $ and the unmatched profile $ \vec{c}_h $. For any full path $ i $ reaching height $ h $ at the midpoint, the total probability and coloring probability decompose as:
\begin{equation}
\begin{aligned}
&p_i = p_{i|h} \cdot p_h = p_{i_L|h} \cdot p_{i_R|h} \cdot p_h, \\
&p_{\vec{c}|i} = p_{\vec{c}|\vec{c}_h,i} \cdot p_{\vec{c}_h} = p_{\vec{c}_L|\vec{c}_h,i_L} \cdot p_{\vec{c}_R|\vec{c}_h,i_R} \cdot p_{\vec{c}_h},
\end{aligned}
\end{equation}
where $ p_h $ is the probability that an uncolored path reaches height $ h $ at the cut $ L/2 $, and $ p_{\vec{c}_h} = 2^{-h} $ is the probability of assigning an unmatched color profile of size $ h $.
Therefore, the Schmidt decomposition of the state is 
\begin{equation}
    |\Psi\rangle = \sum_{h} \sum_{\Vec{c}_h} \sqrt{2^{-h}p_h} |w_{L,h}^{\Vec{c}_h}\rangle \otimes |w_{R,h}^{\Vec{c}_h}\rangle
\end{equation}
Then we find the entanglement entropy,
\begin{equation}
    \begin{aligned}
        S & = - \sum_h p_h \log(2^{-h}p_h)\\
        & = \sum_h hp_h \log2 - \sum_h  p_h \log p_h.\\
    \end{aligned}
\end{equation}
We define $\alpha=\log 2$, $\langle h \rangle = \sum_h hp_h$ as the average height in the classical dynamics, and we let $S_{\text{uncolored}}=-\sum_h  p_h \log p_h$. Then we can simplify the entropy 
\begin{equation}
    S = \alpha \langle h \rangle + S_{\text{uncolored}}.
\end{equation}
In two dimensions, the derivation remains the same, except that the normalization from the number of coloring changes because at the cut the surface $h_x$ has to be matched, i.e.
\begin{equation}
    |\Psi\rangle = \sum_{h_x} \sum_{\Vec{c}_{h_x}} \sqrt{2^{-A_{h_x}}p_{h_x}} |w_{L,h_x}^{\Vec{c}_{h_x}}\rangle \otimes |w_{R,h_x}^{\Vec{c}_{h_x}}\rangle.
\end{equation}
Then the entanglement entropy is 
\begin{equation}
    \begin{aligned}
        S & = \sum_{h_x} A_{h_x}p_{h_x} \log2 - \sum_{h_x}  p_{h_{x}} \log p_{h_x}\\
        & = \alpha \langle A \rangle + S_{\text{uncolored}}.\\
    \end{aligned}
\end{equation}
This was calculated for a special case in \cite{zhang2023coupled}. We emphasize that since $p_h (p_{h_x})$ is independent of the post-selection, $\langle h \rangle (\langle A\rangle)$ is not affected by the post-selection as well.

\section{Explicit form of Parent Hamiltonian and Matrix Elements of $U_n$}
\subsection{Explicit form of Parent Hamiltonian}
\label{sec:expparentH}
Here we provide explicit construction of projectors in \eqref{eq:parentH}, 
\begin{equation}
    \begin{aligned}
        H = &\sum_{i,t} \left(\sum_{k=1}^3\sum_{c=r,b}\sum_{\Vec{S}}\Pi_{i,t}^{k,c,\Vec{S}}\right) + \Pi_{\mathrm{initial}}+ \Pi_{\mathrm{final}} \\
    & + \Pi_{\mathrm{left}} + \Pi_{\mathrm{right}} + \Pi_\text{color} + H_\text{Gauss}. 
    \end{aligned}
\end{equation}
The initial condition projector is
\begin{equation}
    \begin{aligned}
        \Pi_{\mathrm{initial}} = & \frac{1}{L}\sum_{i=1}^{L+1} |\downarrow\rangle_{i-\frac{1}{2},\frac{1}{2}}\langle \downarrow| \\
        & + \frac{1}{L} \sum_{i=1}^L \vcenter{\hbox{
        \tikzset{every picture/.style={line width=0.75pt}} 

\begin{tikzpicture}[x=0.75pt,y=0.75pt,yscale=-0.8,xscale=0.8]

\draw  [dash pattern={on 4.5pt off 4.5pt}]  (20.53,15.07) -- (51.77,46.3) ;
\draw  [dash pattern={on 4.5pt off 4.5pt}]  (51.77,46.3) -- (83,15.07) ;
\draw  [dash pattern={on 4.5pt off 4.5pt}]  (50.77,46.77) -- (82,78) ;
\draw  [dash pattern={on 4.5pt off 4.5pt}]  (19.53,78) -- (50.77,46.77) ;
\draw    (36.15,19.52) -- (36.15,39.85) ;
\draw [shift={(36.15,41.85)}, rotate = 270] [color={rgb, 255:red, 0; green, 0; blue, 0 }  ][line width=0.75]    (8.74,-3.92) .. controls (5.56,-1.84) and (2.65,-0.53) .. (0,0) .. controls (2.65,0.53) and (5.56,1.84) .. (8.74,3.92)   ;
\draw    (67.38,19.52) -- (67.38,39.85) ;
\draw [shift={(67.38,41.85)}, rotate = 270] [color={rgb, 255:red, 0; green, 0; blue, 0 }  ][line width=0.75]    (8.74,-3.92) .. controls (5.56,-1.84) and (2.65,-0.53) .. (0,0) .. controls (2.65,0.53) and (5.56,1.84) .. (8.74,3.92)   ;
\draw    (35.15,48.52) -- (35.15,68.85) ;
\draw [shift={(35.15,70.85)}, rotate = 270] [color={rgb, 255:red, 0; green, 0; blue, 0 }  ][line width=0.75]    (8.74,-3.92) .. controls (5.56,-1.84) and (2.65,-0.53) .. (0,0) .. controls (2.65,0.53) and (5.56,1.84) .. (8.74,3.92)   ;
\draw    (66.38,48.52) -- (66.38,68.85) ;
\draw [shift={(66.38,70.85)}, rotate = 270] [color={rgb, 255:red, 0; green, 0; blue, 0 }  ][line width=0.75]    (8.74,-3.92) .. controls (5.56,-1.84) and (2.65,-0.53) .. (0,0) .. controls (2.65,0.53) and (5.56,1.84) .. (8.74,3.92)   ;
\draw    (13,12) -- (13,81.56) ;
\draw    (89,12.09) -- (97,45.56) ;
\draw    (97,45.56) -- (93.32,63.42) -- (90,79.56) ;
\draw  [dash pattern={on 4.5pt off 4.5pt}]  (152.53,14.72) -- (183.77,45.96) ;
\draw  [dash pattern={on 4.5pt off 4.5pt}]  (183.77,45.96) -- (215,14.72) ;
\draw  [dash pattern={on 4.5pt off 4.5pt}]  (182.77,46.42) -- (214,77.66) ;
\draw  [dash pattern={on 4.5pt off 4.5pt}]  (151.53,77.66) -- (182.77,46.42) ;
\draw    (168.15,19.17) -- (168.15,39.51) ;
\draw [shift={(168.15,41.51)}, rotate = 270] [color={rgb, 255:red, 0; green, 0; blue, 0 }  ][line width=0.75]    (8.74,-3.92) .. controls (5.56,-1.84) and (2.65,-0.53) .. (0,0) .. controls (2.65,0.53) and (5.56,1.84) .. (8.74,3.92)   ;
\draw    (199.38,19.17) -- (199.38,39.51) ;
\draw [shift={(199.38,41.51)}, rotate = 270] [color={rgb, 255:red, 0; green, 0; blue, 0 }  ][line width=0.75]    (8.74,-3.92) .. controls (5.56,-1.84) and (2.65,-0.53) .. (0,0) .. controls (2.65,0.53) and (5.56,1.84) .. (8.74,3.92)   ;
\draw    (167.15,48.17) -- (167.15,68.51) ;
\draw [shift={(167.15,70.51)}, rotate = 270] [color={rgb, 255:red, 0; green, 0; blue, 0 }  ][line width=0.75]    (8.74,-3.92) .. controls (5.56,-1.84) and (2.65,-0.53) .. (0,0) .. controls (2.65,0.53) and (5.56,1.84) .. (8.74,3.92)   ;
\draw    (198.38,48.17) -- (198.38,68.51) ;
\draw [shift={(198.38,70.51)}, rotate = 270] [color={rgb, 255:red, 0; green, 0; blue, 0 }  ][line width=0.75]    (8.74,-3.92) .. controls (5.56,-1.84) and (2.65,-0.53) .. (0,0) .. controls (2.65,0.53) and (5.56,1.84) .. (8.74,3.92)   ;
\draw    (224,11.09) -- (224,80.66) ;
\draw    (147,12.19) -- (138,45.66) ;
\draw    (138,45.66) -- (142.14,63.52) -- (145.88,79.66) ;

\draw (92.8,71.66) node [anchor=north west][inner sep=0.75pt]  [font=\scriptsize]  {$2i-1,1$};

\end{tikzpicture}
        
        }},
    \end{aligned}\label{eq:initial}
\end{equation}
where the first sum enforces the spin pattern and height profile in \eqref{eq:initialfinalspin}, while the second sum ensures that the surface does not dip below the horizon. Similarly, the final condition projector is 
\begin{equation}
    \begin{aligned}
        \Pi_{\mathrm{final}} = & \frac{1}{L}\sum_{i=1}^{L+1}  |\uparrow\rangle_{i-\frac{1}{2},L+\frac{1}{2}}\langle \uparrow| \\
        & + \frac{1}{L} \sum_{i=1}^L \vcenter{\hbox{
        \tikzset{every picture/.style={line width=0.75pt}} 

\begin{tikzpicture}[x=0.75pt,y=0.75pt,yscale=-0.8,xscale=0.8]

\draw  [dash pattern={on 4.5pt off 4.5pt}]  (20.53,15.07) -- (51.77,46.3) ;
\draw  [dash pattern={on 4.5pt off 4.5pt}]  (51.77,46.3) -- (83,15.07) ;
\draw  [dash pattern={on 4.5pt off 4.5pt}]  (50.77,46.77) -- (82,78) ;
\draw  [dash pattern={on 4.5pt off 4.5pt}]  (19.53,78) -- (50.77,46.77) ;
\draw    (34.98,40.55) -- (35.28,22.22) ;
\draw [shift={(35.32,20.22)}, rotate = 90.94] [color={rgb, 255:red, 0; green, 0; blue, 0 }  ][line width=0.75]    (8.74,-3.92) .. controls (5.56,-1.84) and (2.65,-0.53) .. (0,0) .. controls (2.65,0.53) and (5.56,1.84) .. (8.74,3.92)   ;
\draw    (66.22,40.55) -- (66.52,22.22) ;
\draw [shift={(66.55,20.22)}, rotate = 90.94] [color={rgb, 255:red, 0; green, 0; blue, 0 }  ][line width=0.75]    (8.74,-3.92) .. controls (5.56,-1.84) and (2.65,-0.53) .. (0,0) .. controls (2.65,0.53) and (5.56,1.84) .. (8.74,3.92)   ;
\draw    (34.98,72.55) -- (35.28,54.22) ;
\draw [shift={(35.32,52.22)}, rotate = 90.94] [color={rgb, 255:red, 0; green, 0; blue, 0 }  ][line width=0.75]    (8.74,-3.92) .. controls (5.56,-1.84) and (2.65,-0.53) .. (0,0) .. controls (2.65,0.53) and (5.56,1.84) .. (8.74,3.92)   ;
\draw    (66.22,72.55) -- (66.52,54.22) ;
\draw [shift={(66.55,52.22)}, rotate = 90.94] [color={rgb, 255:red, 0; green, 0; blue, 0 }  ][line width=0.75]    (8.74,-3.92) .. controls (5.56,-1.84) and (2.65,-0.53) .. (0,0) .. controls (2.65,0.53) and (5.56,1.84) .. (8.74,3.92)   ;
\draw    (14,14.09) -- (14,83.66) ;
\draw    (89,14.19) -- (97,47.66) ;
\draw    (97,47.66) -- (93.32,65.52) -- (90,81.66) ;
\draw    (226,12.09) -- (226,81.66) ;
\draw    (149,13.19) -- (140,46.66) ;
\draw    (140,46.66) -- (144.14,64.52) -- (147.88,80.66) ;
\draw  [dash pattern={on 4.5pt off 4.5pt}]  (153.53,15.72) -- (184.77,46.96) ;
\draw  [dash pattern={on 4.5pt off 4.5pt}]  (184.77,46.96) -- (216,15.72) ;
\draw  [dash pattern={on 4.5pt off 4.5pt}]  (183.77,47.42) -- (215,78.66) ;
\draw  [dash pattern={on 4.5pt off 4.5pt}]  (152.53,78.66) -- (183.77,47.42) ;
\draw    (167.98,41.21) -- (168.28,22.87) ;
\draw [shift={(168.32,20.87)}, rotate = 90.94] [color={rgb, 255:red, 0; green, 0; blue, 0 }  ][line width=0.75]    (8.74,-3.92) .. controls (5.56,-1.84) and (2.65,-0.53) .. (0,0) .. controls (2.65,0.53) and (5.56,1.84) .. (8.74,3.92)   ;
\draw    (199.22,41.21) -- (199.52,22.87) ;
\draw [shift={(199.55,20.87)}, rotate = 90.94] [color={rgb, 255:red, 0; green, 0; blue, 0 }  ][line width=0.75]    (8.74,-3.92) .. controls (5.56,-1.84) and (2.65,-0.53) .. (0,0) .. controls (2.65,0.53) and (5.56,1.84) .. (8.74,3.92)   ;
\draw    (167.98,73.21) -- (168.28,54.87) ;
\draw [shift={(168.32,52.87)}, rotate = 90.94] [color={rgb, 255:red, 0; green, 0; blue, 0 }  ][line width=0.75]    (8.74,-3.92) .. controls (5.56,-1.84) and (2.65,-0.53) .. (0,0) .. controls (2.65,0.53) and (5.56,1.84) .. (8.74,3.92)   ;
\draw    (199.22,73.21) -- (199.52,54.87) ;
\draw [shift={(199.55,52.87)}, rotate = 90.94] [color={rgb, 255:red, 0; green, 0; blue, 0 }  ][line width=0.75]    (8.74,-3.92) .. controls (5.56,-1.84) and (2.65,-0.53) .. (0,0) .. controls (2.65,0.53) and (5.56,1.84) .. (8.74,3.92)   ;

\draw (93.32,74.92) node [anchor=north west][inner sep=0.75pt]  [font=\scriptsize]  {$2i-1,L$};

\end{tikzpicture}
        
        }} .
    \end{aligned}\label{eq:final}
\end{equation}
which enforces the spin pattern in Eq. \eqref{eq:initialfinalspin} and prevents the surface from rising back to the horizon. The construction of these constraints in \eqref{eq:initial} and \eqref{eq:final} was also discussed in \cite{bravyi2012criticality,movassagh2016supercritical,zhang2017novel}. The left boundary condition projector, enforcing \eqref{eq:leftrightspins}, is 
\begin{equation}
     \Pi_{\mathrm{left}} =\frac{1}{L} \sum_{t=1}^L |\downarrow\rangle_{\frac{1}{2},t-\frac{1}{2}}\langle \downarrow| + |\uparrow\rangle_{\frac{1}{2},t+\frac{1}{2}}\langle \uparrow|,
\end{equation}
and the right boundary condition projector, enforcing \eqref{eq:leftrightspins}, is
\begin{equation}
     \begin{aligned}
         \Pi_{\mathrm{right}} = & \frac{1}{L} \sum_{t=1}^L |\downarrow\rangle_{L-\frac{1}{2},t-\frac{1}{2}}\langle \downarrow| + |\uparrow\rangle_{L-\frac{1}{2},t+\frac{1}{2}}\langle \uparrow|.
     \end{aligned}
\end{equation} 
The energy penalty of Gauss's law violation at all vertices is 
\begin{equation}
    \begin{aligned}
        H_\text{Gauss} = & \sum_{\substack{i+t \text{ odd}\\ 1\leq i,t \leq L}} \left(s_{i-\frac{1}{2},t+\frac{1}{2}}+s_{i-\frac{1}{2},t-\frac{1}{2}}\right.\\
        & \quad \quad \left.-s_{i+\frac{1}{2},t+\frac{1}{2}}-s_{i+\frac{1}{2},t-\frac{1}{2}} \right)^2
    \end{aligned}
\end{equation}
Finally, the color projector on all vertices is 
\begin{equation}
    \begin{aligned}
        \Pi_\text{color} = & \sum_{\substack{i+t \text{ odd}\\ 1\leq i,t \leq L}} \left( 
    \left|
    \vcenter{\hbox{

\tikzset{every picture/.style={line width=0.75pt}} 

\begin{tikzpicture}[x=0.75pt,y=0.75pt,yscale=-0.5,xscale=0.5]

\draw  [dash pattern={on 4.5pt off 4.5pt}]  (20.53,78) -- (51.77,46.77) ;
\draw  [dash pattern={on 4.5pt off 4.5pt}]  (51.77,46.77) -- (83,78) ;
\draw  [dash pattern={on 4.5pt off 4.5pt}]  (50.77,46.3) -- (82,15.07) ;
\draw  [dash pattern={on 4.5pt off 4.5pt}]  (19.53,15.07) -- (50.77,46.3) ;
\draw    (35.98,52.22) -- (36.28,70.55) ;
\draw [shift={(36.32,72.55)}, rotate = 269.06] [color={rgb, 255:red, 0; green, 0; blue, 0 }  ][line width=0.75]    (10.93,-3.29) .. controls (6.95,-1.4) and (3.31,-0.3) .. (0,0) .. controls (3.31,0.3) and (6.95,1.4) .. (10.93,3.29)   ;
\draw    (67.22,52.22) -- (67.52,70.55) ;
\draw [shift={(67.55,72.55)}, rotate = 269.06] [color={rgb, 255:red, 0; green, 0; blue, 0 }  ][line width=0.75]    (10.93,-3.29) .. controls (6.95,-1.4) and (3.31,-0.3) .. (0,0) .. controls (3.31,0.3) and (6.95,1.4) .. (10.93,3.29)   ;
\draw    (35.98,21.52) -- (36.28,39.85) ;
\draw [shift={(36.32,41.85)}, rotate = 269.06] [color={rgb, 255:red, 0; green, 0; blue, 0 }  ][line width=0.75]    (10.93,-3.29) .. controls (6.95,-1.4) and (3.31,-0.3) .. (0,0) .. controls (3.31,0.3) and (6.95,1.4) .. (10.93,3.29)   ;
\draw    (66.22,20.52) -- (66.52,38.85) ;
\draw [shift={(66.55,40.85)}, rotate = 269.06] [color={rgb, 255:red, 0; green, 0; blue, 0 }  ][line width=0.75]    (10.93,-3.29) .. controls (6.95,-1.4) and (3.31,-0.3) .. (0,0) .. controls (3.31,0.3) and (6.95,1.4) .. (10.93,3.29)   ;

\end{tikzpicture}

    }},0
    \right\rangle_{i,t}\left\langle  \vcenter{\hbox{

\tikzset{every picture/.style={line width=0.75pt}} 

\begin{tikzpicture}[x=0.75pt,y=0.75pt,yscale=-0.5,xscale=0.5]

\draw  [dash pattern={on 4.5pt off 4.5pt}]  (20.53,78) -- (51.77,46.77) ;
\draw  [dash pattern={on 4.5pt off 4.5pt}]  (51.77,46.77) -- (83,78) ;
\draw  [dash pattern={on 4.5pt off 4.5pt}]  (50.77,46.3) -- (82,15.07) ;
\draw  [dash pattern={on 4.5pt off 4.5pt}]  (19.53,15.07) -- (50.77,46.3) ;
\draw    (35.98,52.22) -- (36.28,70.55) ;
\draw [shift={(36.32,72.55)}, rotate = 269.06] [color={rgb, 255:red, 0; green, 0; blue, 0 }  ][line width=0.75]    (10.93,-3.29) .. controls (6.95,-1.4) and (3.31,-0.3) .. (0,0) .. controls (3.31,0.3) and (6.95,1.4) .. (10.93,3.29)   ;
\draw    (67.22,52.22) -- (67.52,70.55) ;
\draw [shift={(67.55,72.55)}, rotate = 269.06] [color={rgb, 255:red, 0; green, 0; blue, 0 }  ][line width=0.75]    (10.93,-3.29) .. controls (6.95,-1.4) and (3.31,-0.3) .. (0,0) .. controls (3.31,0.3) and (6.95,1.4) .. (10.93,3.29)   ;
\draw    (35.98,21.52) -- (36.28,39.85) ;
\draw [shift={(36.32,41.85)}, rotate = 269.06] [color={rgb, 255:red, 0; green, 0; blue, 0 }  ][line width=0.75]    (10.93,-3.29) .. controls (6.95,-1.4) and (3.31,-0.3) .. (0,0) .. controls (3.31,0.3) and (6.95,1.4) .. (10.93,3.29)   ;
\draw    (66.22,20.52) -- (66.52,38.85) ;
\draw [shift={(66.55,40.85)}, rotate = 269.06] [color={rgb, 255:red, 0; green, 0; blue, 0 }  ][line width=0.75]    (10.93,-3.29) .. controls (6.95,-1.4) and (3.31,-0.3) .. (0,0) .. controls (3.31,0.3) and (6.95,1.4) .. (10.93,3.29)   ;

\end{tikzpicture}

    }},0\right| \right.\\
    & + \left|
    \vcenter{\hbox{
    \tikzset{every picture/.style={line width=0.75pt}} 

\begin{tikzpicture}[x=0.75pt,y=0.75pt,yscale=-0.5,xscale=0.5]

\draw  [dash pattern={on 4.5pt off 4.5pt}]  (20.53,15.07) -- (51.77,46.3) ;
\draw  [dash pattern={on 4.5pt off 4.5pt}]  (51.77,46.3) -- (83,15.07) ;
\draw  [dash pattern={on 4.5pt off 4.5pt}]  (50.77,46.77) -- (82,78) ;
\draw  [dash pattern={on 4.5pt off 4.5pt}]  (19.53,78) -- (50.77,46.77) ;
\draw    (35.98,40.85) -- (36.28,22.52) ;
\draw [shift={(36.32,20.52)}, rotate = 90.94] [color={rgb, 255:red, 0; green, 0; blue, 0 }  ][line width=0.75]    (10.93,-3.29) .. controls (6.95,-1.4) and (3.31,-0.3) .. (0,0) .. controls (3.31,0.3) and (6.95,1.4) .. (10.93,3.29)   ;
\draw    (67.22,40.85) -- (67.52,22.52) ;
\draw [shift={(67.55,20.52)}, rotate = 90.94] [color={rgb, 255:red, 0; green, 0; blue, 0 }  ][line width=0.75]    (10.93,-3.29) .. controls (6.95,-1.4) and (3.31,-0.3) .. (0,0) .. controls (3.31,0.3) and (6.95,1.4) .. (10.93,3.29)   ;
\draw    (35.98,71.55) -- (36.28,53.22) ;
\draw [shift={(36.32,51.22)}, rotate = 90.94] [color={rgb, 255:red, 0; green, 0; blue, 0 }  ][line width=0.75]    (10.93,-3.29) .. controls (6.95,-1.4) and (3.31,-0.3) .. (0,0) .. controls (3.31,0.3) and (6.95,1.4) .. (10.93,3.29)   ;
\draw    (66.22,72.55) -- (66.52,54.22) ;
\draw [shift={(66.55,52.22)}, rotate = 90.94] [color={rgb, 255:red, 0; green, 0; blue, 0 }  ][line width=0.75]    (10.93,-3.29) .. controls (6.95,-1.4) and (3.31,-0.3) .. (0,0) .. controls (3.31,0.3) and (6.95,1.4) .. (10.93,3.29)   ;

\end{tikzpicture}

    }},0
    \right\rangle_{i,t}\left\langle  \vcenter{\hbox{
    \tikzset{every picture/.style={line width=0.75pt}} 

\begin{tikzpicture}[x=0.75pt,y=0.75pt,yscale=-0.5,xscale=0.5]

\draw  [dash pattern={on 4.5pt off 4.5pt}]  (20.53,15.07) -- (51.77,46.3) ;
\draw  [dash pattern={on 4.5pt off 4.5pt}]  (51.77,46.3) -- (83,15.07) ;
\draw  [dash pattern={on 4.5pt off 4.5pt}]  (50.77,46.77) -- (82,78) ;
\draw  [dash pattern={on 4.5pt off 4.5pt}]  (19.53,78) -- (50.77,46.77) ;
\draw    (35.98,40.85) -- (36.28,22.52) ;
\draw [shift={(36.32,20.52)}, rotate = 90.94] [color={rgb, 255:red, 0; green, 0; blue, 0 }  ][line width=0.75]    (10.93,-3.29) .. controls (6.95,-1.4) and (3.31,-0.3) .. (0,0) .. controls (3.31,0.3) and (6.95,1.4) .. (10.93,3.29)   ;
\draw    (67.22,40.85) -- (67.52,22.52) ;
\draw [shift={(67.55,20.52)}, rotate = 90.94] [color={rgb, 255:red, 0; green, 0; blue, 0 }  ][line width=0.75]    (10.93,-3.29) .. controls (6.95,-1.4) and (3.31,-0.3) .. (0,0) .. controls (3.31,0.3) and (6.95,1.4) .. (10.93,3.29)   ;
\draw    (35.98,71.55) -- (36.28,53.22) ;
\draw [shift={(36.32,51.22)}, rotate = 90.94] [color={rgb, 255:red, 0; green, 0; blue, 0 }  ][line width=0.75]    (10.93,-3.29) .. controls (6.95,-1.4) and (3.31,-0.3) .. (0,0) .. controls (3.31,0.3) and (6.95,1.4) .. (10.93,3.29)   ;
\draw    (66.22,72.55) -- (66.52,54.22) ;
\draw [shift={(66.55,52.22)}, rotate = 90.94] [color={rgb, 255:red, 0; green, 0; blue, 0 }  ][line width=0.75]    (10.93,-3.29) .. controls (6.95,-1.4) and (3.31,-0.3) .. (0,0) .. controls (3.31,0.3) and (6.95,1.4) .. (10.93,3.29)   ;

\end{tikzpicture}

    }},0\right| \\
    & + \sum_{\substack{\text{other spin}\\ \text{config }\vec{S}\\ \text{ with Gauss's Law}}} \left.\sum_{c=r,g} \left|\vec{S},c\right\rangle_{i,t}\left\langle \vec{S},c\right|
    \right)
    \end{aligned}
\end{equation}

Now we explicitly derive $|D_{i,t}^{k,c,\Vec{S}}\rangle$ and $\Pi_{i,t}^{k,c,\Vec{S}}$. Recall that the minimum deformation is the spins and colors around one plaquette. Then the height difference constraint between the plaquettes implies that the eight surrounding spins should be in the following pattern: the top/bottom two horizontal spins should be the same, and the left/right two vertical spins should be opposite. Then, there are only four remaining categories for the boundaries or surroundings where we can perform legal deformation, classified by the top two spins (both $\uparrow$ or $\downarrow$) and bottom spin (both $\uparrow$ or $\downarrow$). These correspond to superposing the trajectory of deposition/evaporation of the middle block columns in the surface growth while keeping the height of the left/right columns unchanged. The height evolution of the site corresponding to the center plaquette locally forms a Motzkin path, and then, with the nonnegative height constraint, the valid trajectory deformations are 
\begin{enumerate}[label=(\alph*)]
    \item $h \to h+2 \to h$ versus $h \to h \to h$;
    \item $h \to h+2 \to h+2$ versus $h \to h \to h+2$; 
    \item $h \to h \to h-2$ versus $h \to h-2 \to h-2$;
\end{enumerate}
as discussed in the one-dimensional cases \cite{bravyi2012criticality,movassagh2016supercritical,zhang2017novel}. 

Therefore, for the plaquette of $h_i(t)$, we can explicitly write out three possible superpositions according to three boundary conditions,
\begin{equation}
    \begin{aligned}
        |D_{i,t}^{1,c,\Vec{S}}\rangle = &\sqrt{p_{\Vec{S},h+2}/2}\,\, \vcenter{\hbox{
        \tikzset{every picture/.style={line width=0.65pt}} 


        }}, \label{eq:D3c}
    \end{aligned}
\end{equation}
where $\Vec{S}=(s_1,s_2,s_3,s_4)$ satisfying $s_1+s_2=s_3+s_4=0$ and $p_{\Vec{S},h}=p_Lp_Rp_Up_L$ records the probability of evolution at the left, right, upper, and lower vertices, following the probability of a single vertex:
\begin{equation}
    P\left[\vcenter{\hbox{

\tikzset{every picture/.style={line width=0.75pt}} 

%

    }}\right]=1.
\end{equation}
Then we can construct the projector for each plaquette with proper normalization,
\begin{equation}
    \Pi_{i,t}^{k,c,\Vec{S}} = I_{i,t}^{k,c,\Vec{S}} - \frac{1}{p_{\Vec{S},h_1}+p_{\Vec{S},h_2}}|D_{i,t}^{k,c,\Vec{S}}\rangle\langle D_{i,t}^{k,c,\Vec{S}}|, \label{eq:updatingPi}
\end{equation}
where $I_{i,t}^{k,c,\Vec{S}}$ is the identity only on the subspace of allowed spins and color configurations around plaquette $(i,t)$, respecting the color $c$ matching and Gauss's law constraint from $\Vec{S}$, and the top and bottom spins of case $k$. Then the local projector $\Pi^{k,c,\Vec{S}}_{i,t}$ would annihilate the states of superposition of all valid trajectories based on the local probability construction of evaporation and deposition.

To obtain the parent Hamiltonian of uncolored states $|\Psi_{\text{abs,uncolored}}\rangle$, we only need to remove the color qutrits in $\Pi_{i,t}^{k,c,\Vec{S}}$ for $\Pi_{i,t}^{k\Vec{S}}$, and to remove the color projector $\Pi_\text{color}$. The Hamiltonian is modified to
\begin{equation}
   \begin{aligned}
        H = & \sum_{i,t} \left(\sum_{k=1}^3\Pi_{i,t}^{k,\Vec{S}}\right) + \Pi_{\mathrm{initial}}+ \Pi_{\mathrm{final}} \\
        & + \Pi_{\mathrm{left}} + \Pi_{\mathrm{right}} + H_\text{Gauss},
   \end{aligned}
\end{equation}
which is also the same as the sequential generation.

\subsection{Matrix Elements of $U_{L(R)}$ and $U_{\mathsf{E(F)},j}$}
\label{sec:Un}
Recall Figure \ref{fig:Unfactorize}, $U_L$ and $U_R$ radiates out the 2nd and $L$th spin qubit based on the neighboring stacks. Explicitly, $U_L$ radiates the 2nd spin based on the relative $\mathsf{F}$/$\mathsf{E}$ positions in the 1st and 2nd stacks,
\begin{align}
    \left\langle 
    \begin{array}{l}
       \mathsf{F}  \\ [-4pt]
        \hspace{0.7em} \mathsf{E} 
    \end{array},\uparrow
    \right| U_L \left|\begin{array}{l}
       \mathsf{F}  \\ [-4pt]
        \hspace{0.7em} \mathsf{E} 
    \end{array},\downarrow\right\rangle &= 1,\\
    \displaybreak[1]
     \left\langle 
    \begin{array}{l}
       \hspace{0.7em}\mathsf{E}  \\ [-4pt]
         \mathsf{F} 
    \end{array},\downarrow
    \right| U_L \left|\begin{array}{l}
       \hspace{0.7em} \mathsf{E}  \\ [-4pt]
        \mathsf{F} 
    \end{array},\downarrow\right\rangle &= 1.
\end{align}
And $U_R$ radiates the $L$th spin based on the relative $\mathsf{E}$/$\mathsf{F}$ positions in the $(L-1)$th and $L$th stacks,
\begin{align}
    \left\langle 
    \begin{array}{l}
       \mathsf{E}  \\ [-4pt]
        \hspace{0.7em} \mathsf{F} 
    \end{array},\downarrow
    \right| U_R  \left|\begin{array}{l}
       \mathsf{E}  \\ [-4pt]
        \hspace{0.7em} \mathsf{F} 
    \end{array},\downarrow\right\rangle &= 1,\\
    \displaybreak[1]
     \left\langle 
    \begin{array}{l}
       \hspace{0.7em}\mathsf{F}  \\ [-4pt]
         \mathsf{E} 
    \end{array},\uparrow
    \right| U_R \left|\begin{array}{l}
       \hspace{0.7em} \mathsf{F}  \\ [-4pt]
        \mathsf{E} 
    \end{array},\downarrow\right\rangle &= 1. 
\end{align}

Based on the discussion of three cases in Section \ref{sec:seq}, we can explicitly write out the matrix elements of $U_{\mathsf{F}(\mathsf{E}),j}$\footnote{Here we only explicitly write out the part of the stacks that are updated.}. The nonzero elements represent the allowed moves, and the exact values are given by the probability of all the events -- no process, deposition, or evaporation -- that contribute to the surface evolution recorded in the stacks change. The matrix elements of $U_{\mathsf{F},j}$ are given by
\begin{align} \label{eq:UF1}
        \left\langle 
    \begin{array}{c}
        \mathsf{F} \\ [-4pt]
       \mathsf{E}c\mathsf{E }\\[-4pt]
        c
    \end{array},\uparrow\uparrow,0
    \right|U_{\mathsf{F},j}\left|\begin{array}{c}
        \mathsf{F} \\ [-4pt]
       \mathsf{E}c\mathsf{E}\\[-4pt]
        c
    \end{array},\downarrow\downarrow,0\right\rangle & = \sqrt{\frac{1+p}{2}},\\ \label{eq:UF2}
    \left\langle 
    \begin{array}{c}
        0 \\ [-4pt]
       \mathsf{E}0\mathsf{E }\\[-4pt]
      \mathsf{F}
    \end{array},\downarrow\downarrow, c
    \right|U_{\mathsf{F},j}\left|\begin{array}{c}
        \mathsf{F} \\ [-4pt]
       \mathsf{E}c\mathsf{E}\\[-4pt]
        c
    \end{array},\downarrow\downarrow,0\right\rangle & = \sqrt{\frac{1-p}{2}},\\ \label{eq:UF3}
    \displaybreak[1]
    \left\langle 
    \begin{array}{c}
        \mathsf{F} \\ [-4pt]
       \mathsf{E}c\mathsf{E }\\[-4pt]
        c
    \end{array},\uparrow\uparrow,c
    \right|U_{\mathsf{F},j}\left|\begin{array}{c}
       0 \\ [-4pt]
       \mathsf{E}0\mathsf{E }\\[-4pt]
        \mathsf{F}
    \end{array},\downarrow\downarrow,0\right\rangle & = \sqrt{p/4},\\ \label{eq:UF4}
    \displaybreak[1]
    \left\langle 
    \begin{array}{c}
       0 \\ [-4pt]
       \mathsf{E}0\mathsf{E }\\[-4pt]
        \mathsf{F}
    \end{array},\downarrow\downarrow,0
    \right|U_{\mathsf{F},j}\left|\begin{array}{c}
       0 \\ [-4pt]
       \mathsf{E}0\mathsf{E }\\[-4pt]
        \mathsf{F}
    \end{array},\downarrow\downarrow,0\right\rangle & = \sqrt{1-p/2},\\ \label{eq:UF5}
    \displaybreak[1]
    \left\langle 
    \begin{array}{l}
    \mathsf{E} \\[-4pt]
    \hspace{0.55em} \mathsf{F} \\[-4pt]
    \hspace{1em} \mathsf{E}
    \end{array},\downarrow\uparrow,0
    \right|U_{\mathsf{F},j}\left|\begin{array}{l}
    \mathsf{E} \\[-4pt]
    \hspace{0.55em} \mathsf{F} \\[-4pt]
    \hspace{1em} \mathsf{E}
    \end{array},\downarrow\downarrow,0\right\rangle & = 1,
    \\ \label{eq:UF6}
    \displaybreak[1]
    \left\langle 
    \begin{array}{l}
    \hspace{1em}\mathsf{E} \\[-4pt]
    \hspace{0.55em} \mathsf{F} \\[-4pt]
    \mathsf{E}
    \end{array},\uparrow\downarrow,0
    \right|U_{\mathsf{F},j}\left|\begin{array}{l}
    \hspace{1em}\mathsf{E} \\[-4pt]
    \hspace{0.55em} \mathsf{F} \\[-4pt]
    \mathsf{E}
    \end{array},\downarrow\downarrow,0\right\rangle & = 1,
\end{align}
where $c=r,g$ splits the probability of uncolored events. And the matrix elements of $U_{\mathsf{E},j}$, which is the same $U_{\mathsf{E},j}$ with the exchange of symbols $\mathsf{E}$ and $\mathsf{F}$, are given by
\begin{align} \label{eq:UE1}
        \left\langle 
    \begin{array}{c}
        \mathsf{E} \\ [-4pt]
       \mathsf{F}c\mathsf{F}\\[-4pt]
        c
    \end{array},\uparrow\uparrow,0
    \right|U_{\mathsf{E},j}\left|\begin{array}{c}
        \mathsf{E} \\ [-4pt]
       \mathsf{F}c\mathsf{F}\\[-4pt]
        c
    \end{array},\downarrow\downarrow,0\right\rangle & = \sqrt{\frac{1+p}{2}},\\ \label{eq:UE2}
    \displaybreak[1]
    \left\langle 
    \begin{array}{c}
        0 \\ [-4pt]
       \mathsf{F}0\mathsf{F }\\[-4pt]
      \mathsf{E}
    \end{array},\downarrow\downarrow, c
    \right|U_{\mathsf{E},j}\left|\begin{array}{c}
        \mathsf{E} \\ [-4pt]
       \mathsf{F}c\mathsf{F}\\[-4pt]
        c
    \end{array},\downarrow\downarrow,0\right\rangle & = \sqrt{\frac{1-p}{2}},\\ \label{eq:UE3}
    \displaybreak[1]
    \left\langle 
    \begin{array}{c}
        \mathsf{E} \\ [-4pt]
       \mathsf{F}c\mathsf{F}\\[-4pt]
        c
    \end{array},\uparrow\uparrow,c
    \right|U_{\mathsf{E},j}\left|\begin{array}{c}
       0 \\ [-4pt]
       \mathsf{F}0\mathsf{F}\\[-4pt]
        \mathsf{E}
    \end{array},\downarrow\downarrow,0\right\rangle & = \sqrt{p/4},\\ \label{eq:UE4}
    \displaybreak[1]
    \left\langle 
    \begin{array}{c}
       0 \\ [-4pt]
       \mathsf{F}0\mathsf{F }\\[-4pt]
        \mathsf{E}
    \end{array},\downarrow\downarrow,0
    \right|U_{\mathsf{E},j}\left|\begin{array}{c}
       0 \\ [-4pt]
       \mathsf{F}0\mathsf{F }\\[-4pt]
        \mathsf{E}
    \end{array},\downarrow\downarrow,0\right\rangle & = \sqrt{1-p/2},\\ \label{eq:UE5}
    \displaybreak[1]
    \left\langle 
    \begin{array}{l}
    \mathsf{F} \\[-4pt]
    \hspace{0.45em} \mathsf{E} \\[-4pt]
    \hspace{1em} \mathsf{F}
    \end{array},\downarrow\uparrow,0
    \right|U_{\mathsf{E},j}\left|\begin{array}{l}
    \mathsf{F} \\[-4pt]
    \hspace{0.45em} \mathsf{E} \\[-4pt]
    \hspace{1em} \mathsf{F}
    \end{array},\downarrow\downarrow,0\right\rangle & = 1,
    \\ \label{eq:UE6}
    \displaybreak[1]
    \left\langle 
    \begin{array}{l}
    \hspace{1em}\mathsf{F} \\[-4pt]
    \hspace{0.5em} \mathsf{E} \\[-4pt]
    \mathsf{F}
    \end{array},\uparrow\downarrow,0
    \right|U_{\mathsf{E},j}\left|\begin{array}{l}
    \hspace{1em}\mathsf{F} \\[-4pt]
    \hspace{0.55em} \mathsf{E} \\[-4pt]
    \mathsf{F}
    \end{array},\downarrow\downarrow,0\right\rangle & = 1.
\end{align}
For each case, the probabilities of all possible moves sum to 1. 
\end{appendix}

\bibliography{reference}

\begin{thebibliography}{67}%
\makeatletter
\providecommand \@ifxundefined [1]{%
 \@ifx{#1\undefined}
}%
\providecommand \@ifnum [1]{%
 \ifnum #1\expandafter \@firstoftwo
 \else \expandafter \@secondoftwo
 \fi
}%
\providecommand \@ifx [1]{%
 \ifx #1\expandafter \@firstoftwo
 \else \expandafter \@secondoftwo
 \fi
}%
\providecommand \natexlab [1]{#1}%
\providecommand \enquote  [1]{``#1''}%
\providecommand \bibnamefont  [1]{#1}%
\providecommand \bibfnamefont [1]{#1}%
\providecommand \citenamefont [1]{#1}%
\providecommand \href@noop [0]{\@secondoftwo}%
\providecommand \href [0]{\begingroup \@sanitize@url \@href}%
\providecommand \@href[1]{\@@startlink{#1}\@@href}%
\providecommand \@@href[1]{\endgroup#1\@@endlink}%
\providecommand \@sanitize@url [0]{\catcode `\\12\catcode `\$12\catcode `\&12\catcode `\#12\catcode `\^12\catcode `\_12\catcode `\%12\relax}%
\providecommand \@@startlink[1]{}%
\providecommand \@@endlink[0]{}%
\providecommand \url  [0]{\begingroup\@sanitize@url \@url }%
\providecommand \@url [1]{\endgroup\@href {#1}{\urlprefix }}%
\providecommand \urlprefix  [0]{URL }%
\providecommand \Eprint [0]{\href }%
\providecommand \doibase [0]{https://doi.org/}%
\providecommand \selectlanguage [0]{\@gobble}%
\providecommand \bibinfo  [0]{\@secondoftwo}%
\providecommand \bibfield  [0]{\@secondoftwo}%
\providecommand \translation [1]{[#1]}%
\providecommand \BibitemOpen [0]{}%
\providecommand \bibitemStop [0]{}%
\providecommand \bibitemNoStop [0]{.\EOS\space}%
\providecommand \EOS [0]{\spacefactor3000\relax}%
\providecommand \BibitemShut  [1]{\csname bibitem#1\endcsname}%
\let\auto@bib@innerbib\@empty
\bibitem [{\citenamefont {Hastings}(2007)}]{Hastings_2007}%
  \BibitemOpen
  \bibfield  {author} {\bibinfo {author} {\bibfnamefont {M.~B.}\ \bibnamefont {Hastings}},\ }\bibfield  {title} {\bibinfo {title} {An area law for one-dimensional quantum systems},\ }\href {https://doi.org/10.1088/1742-5468/2007/08/P08024} {\bibfield  {journal} {\bibinfo  {journal} {Journal of Statistical Mechanics: Theory and Experiment}\ }\textbf {\bibinfo {volume} {2007}},\ \bibinfo {pages} {P08024} (\bibinfo {year} {2007})}\BibitemShut {NoStop}%
\bibitem [{\citenamefont {Anshu}\ \emph {et~al.}(2022)\citenamefont {Anshu}, \citenamefont {Arad},\ and\ \citenamefont {Gosset}}]{10.1145/3519935.3519962}%
  \BibitemOpen
  \bibfield  {author} {\bibinfo {author} {\bibfnamefont {A.}~\bibnamefont {Anshu}}, \bibinfo {author} {\bibfnamefont {I.}~\bibnamefont {Arad}},\ and\ \bibinfo {author} {\bibfnamefont {D.}~\bibnamefont {Gosset}},\ }\bibfield  {title} {\bibinfo {title} {An area law for 2d frustration-free spin systems},\ }in\ \href {https://doi.org/10.1145/3519935.3519962} {\emph {\bibinfo {booktitle} {Proceedings of the 54th Annual ACM SIGACT Symposium on Theory of Computing}}},\ \bibinfo {series and number} {STOC 2022}\ (\bibinfo  {publisher} {Association for Computing Machinery},\ \bibinfo {address} {New York, NY, USA},\ \bibinfo {year} {2022})\ p.\ \bibinfo {pages} {12–18}\BibitemShut {NoStop}%
\bibitem [{\citenamefont {Eisert}\ \emph {et~al.}(2010)\citenamefont {Eisert}, \citenamefont {Cramer},\ and\ \citenamefont {Plenio}}]{RevModPhys.82.277}%
  \BibitemOpen
  \bibfield  {author} {\bibinfo {author} {\bibfnamefont {J.}~\bibnamefont {Eisert}}, \bibinfo {author} {\bibfnamefont {M.}~\bibnamefont {Cramer}},\ and\ \bibinfo {author} {\bibfnamefont {M.~B.}\ \bibnamefont {Plenio}},\ }\bibfield  {title} {\bibinfo {title} {Colloquium: Area laws for the entanglement entropy},\ }\href {https://doi.org/10.1103/RevModPhys.82.277} {\bibfield  {journal} {\bibinfo  {journal} {Rev. Mod. Phys.}\ }\textbf {\bibinfo {volume} {82}},\ \bibinfo {pages} {277} (\bibinfo {year} {2010})}\BibitemShut {NoStop}%
\bibitem [{\citenamefont {Levin}\ and\ \citenamefont {Wen}(2006)}]{PhysRevLett.96.110405}%
  \BibitemOpen
  \bibfield  {author} {\bibinfo {author} {\bibfnamefont {M.}~\bibnamefont {Levin}}\ and\ \bibinfo {author} {\bibfnamefont {X.-G.}\ \bibnamefont {Wen}},\ }\bibfield  {title} {\bibinfo {title} {Detecting topological order in a ground state wave function},\ }\href {https://doi.org/10.1103/PhysRevLett.96.110405} {\bibfield  {journal} {\bibinfo  {journal} {Phys. Rev. Lett.}\ }\textbf {\bibinfo {volume} {96}},\ \bibinfo {pages} {110405} (\bibinfo {year} {2006})}\BibitemShut {NoStop}%
\bibitem [{\citenamefont {Kitaev}\ and\ \citenamefont {Preskill}(2006)}]{PhysRevLett.96.110404}%
  \BibitemOpen
  \bibfield  {author} {\bibinfo {author} {\bibfnamefont {A.}~\bibnamefont {Kitaev}}\ and\ \bibinfo {author} {\bibfnamefont {J.}~\bibnamefont {Preskill}},\ }\bibfield  {title} {\bibinfo {title} {Topological entanglement entropy},\ }\href {https://doi.org/10.1103/PhysRevLett.96.110404} {\bibfield  {journal} {\bibinfo  {journal} {Phys. Rev. Lett.}\ }\textbf {\bibinfo {volume} {96}},\ \bibinfo {pages} {110404} (\bibinfo {year} {2006})}\BibitemShut {NoStop}%
\bibitem [{Note1()}]{Note1}%
  \BibitemOpen
  \bibinfo {note} {This means the entanglement entropy of a subregion $D$ of the state scales faster than the size of $\partial D$.}\BibitemShut {Stop}%
\bibitem [{\citenamefont {Calabrese}\ and\ \citenamefont {Cardy}(2004)}]{PasqualeCalabrese_2004}%
  \BibitemOpen
  \bibfield  {author} {\bibinfo {author} {\bibfnamefont {P.}~\bibnamefont {Calabrese}}\ and\ \bibinfo {author} {\bibfnamefont {J.}~\bibnamefont {Cardy}},\ }\bibfield  {title} {\bibinfo {title} {Entanglement entropy and quantum field theory},\ }\href {https://doi.org/10.1088/1742-5468/2004/06/P06002} {\bibfield  {journal} {\bibinfo  {journal} {Journal of Statistical Mechanics: Theory and Experiment}\ }\textbf {\bibinfo {volume} {2004}},\ \bibinfo {pages} {P06002} (\bibinfo {year} {2004})}\BibitemShut {NoStop}%
\bibitem [{\citenamefont {Wolf}(2006)}]{PhysRevLett.96.010404}%
  \BibitemOpen
  \bibfield  {author} {\bibinfo {author} {\bibfnamefont {M.~M.}\ \bibnamefont {Wolf}},\ }\bibfield  {title} {\bibinfo {title} {Violation of the entropic area law for fermions},\ }\href {https://doi.org/10.1103/PhysRevLett.96.010404} {\bibfield  {journal} {\bibinfo  {journal} {Phys. Rev. Lett.}\ }\textbf {\bibinfo {volume} {96}},\ \bibinfo {pages} {010404} (\bibinfo {year} {2006})}\BibitemShut {NoStop}%
\bibitem [{\citenamefont {Gioev}\ and\ \citenamefont {Klich}(2006)}]{PhysRevLett.96.100503}%
  \BibitemOpen
  \bibfield  {author} {\bibinfo {author} {\bibfnamefont {D.}~\bibnamefont {Gioev}}\ and\ \bibinfo {author} {\bibfnamefont {I.}~\bibnamefont {Klich}},\ }\bibfield  {title} {\bibinfo {title} {Entanglement entropy of fermions in any dimension and the widom conjecture},\ }\href {https://doi.org/10.1103/PhysRevLett.96.100503} {\bibfield  {journal} {\bibinfo  {journal} {Phys. Rev. Lett.}\ }\textbf {\bibinfo {volume} {96}},\ \bibinfo {pages} {100503} (\bibinfo {year} {2006})}\BibitemShut {NoStop}%
\bibitem [{\citenamefont {Bravyi}\ \emph {et~al.}(2012)\citenamefont {Bravyi}, \citenamefont {Caha}, \citenamefont {Movassagh}, \citenamefont {Nagaj},\ and\ \citenamefont {Shor}}]{bravyi2012criticality}%
  \BibitemOpen
  \bibfield  {author} {\bibinfo {author} {\bibfnamefont {S.}~\bibnamefont {Bravyi}}, \bibinfo {author} {\bibfnamefont {L.}~\bibnamefont {Caha}}, \bibinfo {author} {\bibfnamefont {R.}~\bibnamefont {Movassagh}}, \bibinfo {author} {\bibfnamefont {D.}~\bibnamefont {Nagaj}},\ and\ \bibinfo {author} {\bibfnamefont {P.~W.}\ \bibnamefont {Shor}},\ }\bibfield  {title} {\bibinfo {title} {Criticality without frustration for quantum spin-1 chains},\ }\href {https://doi.org/10.1103/PhysRevLett.109.207202} {\bibfield  {journal} {\bibinfo  {journal} {Physical review letters}\ }\textbf {\bibinfo {volume} {109}},\ \bibinfo {pages} {207202} (\bibinfo {year} {2012})}\BibitemShut {NoStop}%
\bibitem [{\citenamefont {Movassagh}\ and\ \citenamefont {Shor}(2016)}]{movassagh2016supercritical}%
  \BibitemOpen
  \bibfield  {author} {\bibinfo {author} {\bibfnamefont {R.}~\bibnamefont {Movassagh}}\ and\ \bibinfo {author} {\bibfnamefont {P.~W.}\ \bibnamefont {Shor}},\ }\bibfield  {title} {\bibinfo {title} {Supercritical entanglement in local systems: Counterexample to the area law for quantum matter},\ }\href {https://doi.org/10.1073/pnas.1605716113} {\bibfield  {journal} {\bibinfo  {journal} {Proceedings of the National Academy of Sciences}\ }\textbf {\bibinfo {volume} {113}},\ \bibinfo {pages} {13278} (\bibinfo {year} {2016})}\BibitemShut {NoStop}%
\bibitem [{\citenamefont {Zhang}\ \emph {et~al.}(2017)\citenamefont {Zhang}, \citenamefont {Ahmadain},\ and\ \citenamefont {Klich}}]{zhang2017novel}%
  \BibitemOpen
  \bibfield  {author} {\bibinfo {author} {\bibfnamefont {Z.}~\bibnamefont {Zhang}}, \bibinfo {author} {\bibfnamefont {A.}~\bibnamefont {Ahmadain}},\ and\ \bibinfo {author} {\bibfnamefont {I.}~\bibnamefont {Klich}},\ }\bibfield  {title} {\bibinfo {title} {Novel quantum phase transition from bounded to extensive entanglement},\ }\href {https://doi.org/10.1073/pnas.1702029114} {\bibfield  {journal} {\bibinfo  {journal} {Proceedings of the National Academy of Sciences}\ }\textbf {\bibinfo {volume} {114}},\ \bibinfo {pages} {5142} (\bibinfo {year} {2017})}\BibitemShut {NoStop}%
\bibitem [{\citenamefont {Alexander}\ \emph {et~al.}(2021)\citenamefont {Alexander}, \citenamefont {Evenbly},\ and\ \citenamefont {Klich}}]{Alexander2021}%
  \BibitemOpen
  \bibfield  {author} {\bibinfo {author} {\bibfnamefont {R.~N.}\ \bibnamefont {Alexander}}, \bibinfo {author} {\bibfnamefont {G.}~\bibnamefont {Evenbly}},\ and\ \bibinfo {author} {\bibfnamefont {I.}~\bibnamefont {Klich}},\ }\bibfield  {title} {\bibinfo {title} {Exact holographic tensor networks for the motzkin spin chain},\ }\href {https://doi.org/10.22331/q-2021-10-12-546} {\bibfield  {journal} {\bibinfo  {journal} {Quantum}\ }\textbf {\bibinfo {volume} {5}},\ \bibinfo {pages} {546} (\bibinfo {year} {2021})}\BibitemShut {NoStop}%
\bibitem [{\citenamefont {Salberger}\ and\ \citenamefont {Korepin}(2018)}]{salberger2018fredkin}%
  \BibitemOpen
  \bibfield  {author} {\bibinfo {author} {\bibfnamefont {O.}~\bibnamefont {Salberger}}\ and\ \bibinfo {author} {\bibfnamefont {V.}~\bibnamefont {Korepin}},\ }\bibfield  {title} {\bibinfo {title} {Fredkin spin chain},\ }in\ \href {https://doi.org/10.1142/9789813233867_0022} {\emph {\bibinfo {booktitle} {Ludwig Faddeev Memorial Volume: A Life In Mathematical Physics}}}\ (\bibinfo  {publisher} {World Scientific},\ \bibinfo {year} {2018})\ pp.\ \bibinfo {pages} {439--458}\BibitemShut {NoStop}%
\bibitem [{\citenamefont {Salberger}\ \emph {et~al.}(2017)\citenamefont {Salberger}, \citenamefont {Udagawa}, \citenamefont {Zhang}, \citenamefont {Katsura}, \citenamefont {Klich},\ and\ \citenamefont {Korepin}}]{Salberger2017}%
  \BibitemOpen
  \bibfield  {author} {\bibinfo {author} {\bibfnamefont {O.}~\bibnamefont {Salberger}}, \bibinfo {author} {\bibfnamefont {T.}~\bibnamefont {Udagawa}}, \bibinfo {author} {\bibfnamefont {Z.}~\bibnamefont {Zhang}}, \bibinfo {author} {\bibfnamefont {H.}~\bibnamefont {Katsura}}, \bibinfo {author} {\bibfnamefont {I.}~\bibnamefont {Klich}},\ and\ \bibinfo {author} {\bibfnamefont {V.}~\bibnamefont {Korepin}},\ }\bibfield  {title} {\bibinfo {title} {Deformed fredkin spin chain with extensive entanglement},\ }\href {https://doi.org/10.1088/1742-5468/aa7143} {\bibfield  {journal} {\bibinfo  {journal} {Journal of Statistical Mechanics: Theory and Experiment}\ }\textbf {\bibinfo {volume} {2017}},\ \bibinfo {pages} {063103} (\bibinfo {year} {2017})}\BibitemShut {NoStop}%
\bibitem [{\citenamefont {Gopalakrishnan}(2025)}]{gopalakrishnan2023push}%
  \BibitemOpen
  \bibfield  {author} {\bibinfo {author} {\bibfnamefont {S.}~\bibnamefont {Gopalakrishnan}},\ }\bibfield  {title} {\bibinfo {title} {Push-down automata as sequential generators of highly entangled states},\ }\href {https://doi.org/10.1088/1751-8121/adab9f} {\bibfield  {journal} {\bibinfo  {journal} {Journal of Physics A: Mathematical and Theoretical}\ }\textbf {\bibinfo {volume} {58}},\ \bibinfo {pages} {055301} (\bibinfo {year} {2025})}\BibitemShut {NoStop}%
\bibitem [{\citenamefont {Crosswhite}\ and\ \citenamefont {Bacon}(2008)}]{crosswhite2008finite}%
  \BibitemOpen
  \bibfield  {author} {\bibinfo {author} {\bibfnamefont {G.~M.}\ \bibnamefont {Crosswhite}}\ and\ \bibinfo {author} {\bibfnamefont {D.}~\bibnamefont {Bacon}},\ }\bibfield  {title} {\bibinfo {title} {Finite automata for caching in matrix product algorithms},\ }\href {https://doi.org/10.1103/PhysRevA.78.012356} {\bibfield  {journal} {\bibinfo  {journal} {Physical Review A—Atomic, Molecular, and Optical Physics}\ }\textbf {\bibinfo {volume} {78}},\ \bibinfo {pages} {012356} (\bibinfo {year} {2008})}\BibitemShut {NoStop}%
\bibitem [{\citenamefont {Florido-Llin{\`a}s}\ \emph {et~al.}(2024)\citenamefont {Florido-Llin{\`a}s}, \citenamefont {Alhambra}, \citenamefont {P{\'e}rez-Garc{\'\i}a},\ and\ \citenamefont {Cirac}}]{florido2024regular}%
  \BibitemOpen
  \bibfield  {author} {\bibinfo {author} {\bibfnamefont {M.}~\bibnamefont {Florido-Llin{\`a}s}}, \bibinfo {author} {\bibfnamefont {{\'A}.~M.}\ \bibnamefont {Alhambra}}, \bibinfo {author} {\bibfnamefont {D.}~\bibnamefont {P{\'e}rez-Garc{\'\i}a}},\ and\ \bibinfo {author} {\bibfnamefont {J.~I.}\ \bibnamefont {Cirac}},\ }\bibfield  {title} {\bibinfo {title} {Regular language quantum states},\ }\href@noop {} {\bibfield  {journal} {\bibinfo  {journal} {arXiv preprint arXiv:2407.17641}\ } (\bibinfo {year} {2024})}\BibitemShut {NoStop}%
\bibitem [{\citenamefont {Calabrese}\ and\ \citenamefont {Cardy}(2005)}]{Calabrese_2005}%
  \BibitemOpen
  \bibfield  {author} {\bibinfo {author} {\bibfnamefont {P.}~\bibnamefont {Calabrese}}\ and\ \bibinfo {author} {\bibfnamefont {J.}~\bibnamefont {Cardy}},\ }\bibfield  {title} {\bibinfo {title} {Evolution of entanglement entropy in one-dimensional systems},\ }\href {https://doi.org/10.1088/1742-5468/2005/04/P04010} {\bibfield  {journal} {\bibinfo  {journal} {Journal of Statistical Mechanics: Theory and Experiment}\ }\textbf {\bibinfo {volume} {2005}},\ \bibinfo {pages} {P04010} (\bibinfo {year} {2005})}\BibitemShut {NoStop}%
\bibitem [{\citenamefont {Nahum}\ \emph {et~al.}(2017)\citenamefont {Nahum}, \citenamefont {Ruhman}, \citenamefont {Vijay},\ and\ \citenamefont {Haah}}]{PhysRevX.7.031016}%
  \BibitemOpen
  \bibfield  {author} {\bibinfo {author} {\bibfnamefont {A.}~\bibnamefont {Nahum}}, \bibinfo {author} {\bibfnamefont {J.}~\bibnamefont {Ruhman}}, \bibinfo {author} {\bibfnamefont {S.}~\bibnamefont {Vijay}},\ and\ \bibinfo {author} {\bibfnamefont {J.}~\bibnamefont {Haah}},\ }\bibfield  {title} {\bibinfo {title} {Quantum entanglement growth under random unitary dynamics},\ }\href {https://doi.org/10.1103/PhysRevX.7.031016} {\bibfield  {journal} {\bibinfo  {journal} {Phys. Rev. X}\ }\textbf {\bibinfo {volume} {7}},\ \bibinfo {pages} {031016} (\bibinfo {year} {2017})}\BibitemShut {NoStop}%
\bibitem [{\citenamefont {Anand}\ \emph {et~al.}(2023{\natexlab{a}})\citenamefont {Anand}, \citenamefont {Hauschild}, \citenamefont {Zhang}, \citenamefont {Potter},\ and\ \citenamefont {Zaletel}}]{PRXQuantum.4.030334}%
  \BibitemOpen
  \bibfield  {author} {\bibinfo {author} {\bibfnamefont {S.}~\bibnamefont {Anand}}, \bibinfo {author} {\bibfnamefont {J.}~\bibnamefont {Hauschild}}, \bibinfo {author} {\bibfnamefont {Y.}~\bibnamefont {Zhang}}, \bibinfo {author} {\bibfnamefont {A.~C.}\ \bibnamefont {Potter}},\ and\ \bibinfo {author} {\bibfnamefont {M.~P.}\ \bibnamefont {Zaletel}},\ }\bibfield  {title} {\bibinfo {title} {Holographic quantum simulation of entanglement renormalization circuits},\ }\href {https://doi.org/10.1103/PRXQuantum.4.030334} {\bibfield  {journal} {\bibinfo  {journal} {PRX Quantum}\ }\textbf {\bibinfo {volume} {4}},\ \bibinfo {pages} {030334} (\bibinfo {year} {2023}{\natexlab{a}})}\BibitemShut {NoStop}%
\bibitem [{\citenamefont {Lu}\ \emph {et~al.}(2022)\citenamefont {Lu}, \citenamefont {Lessa}, \citenamefont {Kim},\ and\ \citenamefont {Hsieh}}]{PRXQuantum.3.040337}%
  \BibitemOpen
  \bibfield  {author} {\bibinfo {author} {\bibfnamefont {T.-C.}\ \bibnamefont {Lu}}, \bibinfo {author} {\bibfnamefont {L.~A.}\ \bibnamefont {Lessa}}, \bibinfo {author} {\bibfnamefont {I.~H.}\ \bibnamefont {Kim}},\ and\ \bibinfo {author} {\bibfnamefont {T.~H.}\ \bibnamefont {Hsieh}},\ }\bibfield  {title} {\bibinfo {title} {Measurement as a shortcut to long-range entangled quantum matter},\ }\href {https://doi.org/10.1103/PRXQuantum.3.040337} {\bibfield  {journal} {\bibinfo  {journal} {PRX Quantum}\ }\textbf {\bibinfo {volume} {3}},\ \bibinfo {pages} {040337} (\bibinfo {year} {2022})}\BibitemShut {NoStop}%
\bibitem [{\citenamefont {Zhu}\ \emph {et~al.}(2023)\citenamefont {Zhu}, \citenamefont {Tantivasadakarn}, \citenamefont {Vishwanath}, \citenamefont {Trebst},\ and\ \citenamefont {Verresen}}]{zhu2023nishimori}%
  \BibitemOpen
  \bibfield  {author} {\bibinfo {author} {\bibfnamefont {G.-Y.}\ \bibnamefont {Zhu}}, \bibinfo {author} {\bibfnamefont {N.}~\bibnamefont {Tantivasadakarn}}, \bibinfo {author} {\bibfnamefont {A.}~\bibnamefont {Vishwanath}}, \bibinfo {author} {\bibfnamefont {S.}~\bibnamefont {Trebst}},\ and\ \bibinfo {author} {\bibfnamefont {R.}~\bibnamefont {Verresen}},\ }\bibfield  {title} {\bibinfo {title} {Nishimori’s cat: stable long-range entanglement from finite-depth unitaries and weak measurements},\ }\href {https://doi.org/10.1103/PhysRevLett.131.200201} {\bibfield  {journal} {\bibinfo  {journal} {Physical Review Letters}\ }\textbf {\bibinfo {volume} {131}},\ \bibinfo {pages} {200201} (\bibinfo {year} {2023})}\BibitemShut {NoStop}%
\bibitem [{\citenamefont {Foss-Feig}\ \emph {et~al.}(2023)\citenamefont {Foss-Feig}, \citenamefont {Tikku}, \citenamefont {Lu}, \citenamefont {Mayer}, \citenamefont {Iqbal}, \citenamefont {Gatterman}, \citenamefont {Gerber}, \citenamefont {Gilmore}, \citenamefont {Gresh}, \citenamefont {Hankin} \emph {et~al.}}]{foss2023experimental}%
  \BibitemOpen
  \bibfield  {author} {\bibinfo {author} {\bibfnamefont {M.}~\bibnamefont {Foss-Feig}}, \bibinfo {author} {\bibfnamefont {A.}~\bibnamefont {Tikku}}, \bibinfo {author} {\bibfnamefont {T.-C.}\ \bibnamefont {Lu}}, \bibinfo {author} {\bibfnamefont {K.}~\bibnamefont {Mayer}}, \bibinfo {author} {\bibfnamefont {M.}~\bibnamefont {Iqbal}}, \bibinfo {author} {\bibfnamefont {T.~M.}\ \bibnamefont {Gatterman}}, \bibinfo {author} {\bibfnamefont {J.~A.}\ \bibnamefont {Gerber}}, \bibinfo {author} {\bibfnamefont {K.}~\bibnamefont {Gilmore}}, \bibinfo {author} {\bibfnamefont {D.}~\bibnamefont {Gresh}}, \bibinfo {author} {\bibfnamefont {A.}~\bibnamefont {Hankin}}, \emph {et~al.},\ }\bibfield  {title} {\bibinfo {title} {Experimental demonstration of the advantage of adaptive quantum circuits},\ }\href@noop {} {\bibfield  {journal} {\bibinfo  {journal} {arXiv preprint arXiv:2302.03029}\ } (\bibinfo {year} {2023})}\BibitemShut {NoStop}%
\bibitem [{\citenamefont {Iqbal}\ \emph {et~al.}(2024)\citenamefont {Iqbal}, \citenamefont {Tantivasadakarn}, \citenamefont {Gatterman}, \citenamefont {Gerber}, \citenamefont {Gilmore}, \citenamefont {Gresh}, \citenamefont {Hankin}, \citenamefont {Hewitt}, \citenamefont {Horst}, \citenamefont {Matheny} \emph {et~al.}}]{iqbal2024topological}%
  \BibitemOpen
  \bibfield  {author} {\bibinfo {author} {\bibfnamefont {M.}~\bibnamefont {Iqbal}}, \bibinfo {author} {\bibfnamefont {N.}~\bibnamefont {Tantivasadakarn}}, \bibinfo {author} {\bibfnamefont {T.~M.}\ \bibnamefont {Gatterman}}, \bibinfo {author} {\bibfnamefont {J.~A.}\ \bibnamefont {Gerber}}, \bibinfo {author} {\bibfnamefont {K.}~\bibnamefont {Gilmore}}, \bibinfo {author} {\bibfnamefont {D.}~\bibnamefont {Gresh}}, \bibinfo {author} {\bibfnamefont {A.}~\bibnamefont {Hankin}}, \bibinfo {author} {\bibfnamefont {N.}~\bibnamefont {Hewitt}}, \bibinfo {author} {\bibfnamefont {C.~V.}\ \bibnamefont {Horst}}, \bibinfo {author} {\bibfnamefont {M.}~\bibnamefont {Matheny}}, \emph {et~al.},\ }\bibfield  {title} {\bibinfo {title} {Topological order from measurements and feed-forward on a trapped ion quantum computer},\ }\href {https://doi.org/10.1038/s42005-024-01698-3} {\bibfield  {journal} {\bibinfo  {journal} {Communications Physics}\ }\textbf {\bibinfo {volume} {7}},\ \bibinfo {pages} {205} (\bibinfo {year}
  {2024})}\BibitemShut {NoStop}%
\bibitem [{\citenamefont {Sch\"on}\ \emph {et~al.}(2005)\citenamefont {Sch\"on}, \citenamefont {Solano}, \citenamefont {Verstraete}, \citenamefont {Cirac},\ and\ \citenamefont {Wolf}}]{Schon2005}%
  \BibitemOpen
  \bibfield  {author} {\bibinfo {author} {\bibfnamefont {C.}~\bibnamefont {Sch\"on}}, \bibinfo {author} {\bibfnamefont {E.}~\bibnamefont {Solano}}, \bibinfo {author} {\bibfnamefont {F.}~\bibnamefont {Verstraete}}, \bibinfo {author} {\bibfnamefont {J.~I.}\ \bibnamefont {Cirac}},\ and\ \bibinfo {author} {\bibfnamefont {M.~M.}\ \bibnamefont {Wolf}},\ }\bibfield  {title} {\bibinfo {title} {Sequential generation of entangled multiqubit states},\ }\href {https://doi.org/10.1103/PhysRevLett.95.110503} {\bibfield  {journal} {\bibinfo  {journal} {Phys. Rev. Lett.}\ }\textbf {\bibinfo {volume} {95}},\ \bibinfo {pages} {110503} (\bibinfo {year} {2005})}\BibitemShut {NoStop}%
\bibitem [{\citenamefont {Bañuls}\ \emph {et~al.}(2008)\citenamefont {Bañuls}, \citenamefont {Pérez-García}, \citenamefont {Wolf}, \citenamefont {Verstraete},\ and\ \citenamefont {Cirac}}]{Banuls2008}%
  \BibitemOpen
  \bibfield  {author} {\bibinfo {author} {\bibfnamefont {M.~C.}\ \bibnamefont {Bañuls}}, \bibinfo {author} {\bibfnamefont {D.}~\bibnamefont {Pérez-García}}, \bibinfo {author} {\bibfnamefont {M.~M.}\ \bibnamefont {Wolf}}, \bibinfo {author} {\bibfnamefont {F.}~\bibnamefont {Verstraete}},\ and\ \bibinfo {author} {\bibfnamefont {J.~I.}\ \bibnamefont {Cirac}},\ }\bibfield  {title} {\bibinfo {title} {Sequentially generated states for the study of two-dimensional systems},\ }\href {https://doi.org/10.1103/PhysRevA.77.052306} {\bibfield  {journal} {\bibinfo  {journal} {Phys. Rev. A}\ }\textbf {\bibinfo {volume} {77}},\ \bibinfo {pages} {052306} (\bibinfo {year} {2008})}\BibitemShut {NoStop}%
\bibitem [{\citenamefont {Wei}\ \emph {et~al.}(2022)\citenamefont {Wei}, \citenamefont {Malz},\ and\ \citenamefont {Cirac}}]{Wei2022}%
  \BibitemOpen
  \bibfield  {author} {\bibinfo {author} {\bibfnamefont {Z.-Y.}\ \bibnamefont {Wei}}, \bibinfo {author} {\bibfnamefont {D.}~\bibnamefont {Malz}},\ and\ \bibinfo {author} {\bibfnamefont {J.~I.}\ \bibnamefont {Cirac}},\ }\bibfield  {title} {\bibinfo {title} {Sequential generation of projected entangled-pair states},\ }\href {https://doi.org/10.1103/PhysRevLett.128.010607} {\bibfield  {journal} {\bibinfo  {journal} {Phys. Rev. Lett.}\ }\textbf {\bibinfo {volume} {128}},\ \bibinfo {pages} {010607} (\bibinfo {year} {2022})}\BibitemShut {NoStop}%
\bibitem [{\citenamefont {Osborne}\ \emph {et~al.}(2010)\citenamefont {Osborne}, \citenamefont {Eisert},\ and\ \citenamefont {Verstraete}}]{Osborne2010}%
  \BibitemOpen
  \bibfield  {author} {\bibinfo {author} {\bibfnamefont {T.~J.}\ \bibnamefont {Osborne}}, \bibinfo {author} {\bibfnamefont {J.}~\bibnamefont {Eisert}},\ and\ \bibinfo {author} {\bibfnamefont {F.}~\bibnamefont {Verstraete}},\ }\bibfield  {title} {\bibinfo {title} {Holographic quantum states},\ }\href {https://doi.org/10.1103/PhysRevLett.105.260401} {\bibfield  {journal} {\bibinfo  {journal} {Phys. Rev. Lett.}\ }\textbf {\bibinfo {volume} {105}},\ \bibinfo {pages} {260401} (\bibinfo {year} {2010})}\BibitemShut {NoStop}%
\bibitem [{\citenamefont {Wang}\ \emph {et~al.}(2017)\citenamefont {Wang}, \citenamefont {Stephen},\ and\ \citenamefont {Raussendorf}}]{Wang2017}%
  \BibitemOpen
  \bibfield  {author} {\bibinfo {author} {\bibfnamefont {D.-S.}\ \bibnamefont {Wang}}, \bibinfo {author} {\bibfnamefont {D.~T.}\ \bibnamefont {Stephen}},\ and\ \bibinfo {author} {\bibfnamefont {R.}~\bibnamefont {Raussendorf}},\ }\bibfield  {title} {\bibinfo {title} {Qudit quantum computation on matrix product states with global symmetry},\ }\href {https://doi.org/10.1103/PhysRevA.95.032312} {\bibfield  {journal} {\bibinfo  {journal} {Phys. Rev. A}\ }\textbf {\bibinfo {volume} {95}},\ \bibinfo {pages} {032312} (\bibinfo {year} {2017})}\BibitemShut {NoStop}%
\bibitem [{\citenamefont {Astrakhantsev}\ \emph {et~al.}(2022)\citenamefont {Astrakhantsev}, \citenamefont {Lin}, \citenamefont {Pollmann},\ and\ \citenamefont {Smith}}]{Astrakhantsev2022}%
  \BibitemOpen
  \bibfield  {author} {\bibinfo {author} {\bibfnamefont {N.}~\bibnamefont {Astrakhantsev}}, \bibinfo {author} {\bibfnamefont {S.-H.}\ \bibnamefont {Lin}}, \bibinfo {author} {\bibfnamefont {F.}~\bibnamefont {Pollmann}},\ and\ \bibinfo {author} {\bibfnamefont {A.}~\bibnamefont {Smith}},\ }\bibfield  {title} {\bibinfo {title} {Time evolution of uniform sequential circuits},\ }\href {https://arxiv.org/abs/2210.03751} {\bibfield  {journal} {\bibinfo  {journal} {arXiv preprint arXiv:2210.03751}\ } (\bibinfo {year} {2022})}\BibitemShut {NoStop}%
\bibitem [{\citenamefont {Barratt}\ \emph {et~al.}(2021)\citenamefont {Barratt}, \citenamefont {Dborin}, \citenamefont {Bal}, \citenamefont {Stojevic}, \citenamefont {Pollmann},\ and\ \citenamefont {Green}}]{Barratt2021}%
  \BibitemOpen
  \bibfield  {author} {\bibinfo {author} {\bibfnamefont {F.}~\bibnamefont {Barratt}}, \bibinfo {author} {\bibfnamefont {J.}~\bibnamefont {Dborin}}, \bibinfo {author} {\bibfnamefont {M.}~\bibnamefont {Bal}}, \bibinfo {author} {\bibfnamefont {V.}~\bibnamefont {Stojevic}}, \bibinfo {author} {\bibfnamefont {F.}~\bibnamefont {Pollmann}},\ and\ \bibinfo {author} {\bibfnamefont {A.~G.}\ \bibnamefont {Green}},\ }\bibfield  {title} {\bibinfo {title} {Parallel quantum simulation of large systems on small nisq computers},\ }\href {https://doi.org/10.1038/s41534-021-00414-9} {\bibfield  {journal} {\bibinfo  {journal} {npj Quantum Information}\ }\textbf {\bibinfo {volume} {7}},\ \bibinfo {pages} {79} (\bibinfo {year} {2021})}\BibitemShut {NoStop}%
\bibitem [{\citenamefont {Foss-Feig}\ \emph {et~al.}(2021)\citenamefont {Foss-Feig}, \citenamefont {Hayes}, \citenamefont {Dreiling}, \citenamefont {Figgatt}, \citenamefont {Gaebler}, \citenamefont {Moses}, \citenamefont {Pino},\ and\ \citenamefont {Potter}}]{FossFeig2021}%
  \BibitemOpen
  \bibfield  {author} {\bibinfo {author} {\bibfnamefont {M.}~\bibnamefont {Foss-Feig}}, \bibinfo {author} {\bibfnamefont {D.}~\bibnamefont {Hayes}}, \bibinfo {author} {\bibfnamefont {J.~M.}\ \bibnamefont {Dreiling}}, \bibinfo {author} {\bibfnamefont {C.}~\bibnamefont {Figgatt}}, \bibinfo {author} {\bibfnamefont {J.~P.}\ \bibnamefont {Gaebler}}, \bibinfo {author} {\bibfnamefont {S.~A.}\ \bibnamefont {Moses}}, \bibinfo {author} {\bibfnamefont {J.~M.}\ \bibnamefont {Pino}},\ and\ \bibinfo {author} {\bibfnamefont {A.~C.}\ \bibnamefont {Potter}},\ }\bibfield  {title} {\bibinfo {title} {Holographic quantum algorithms for simulating correlated spin systems},\ }\href {https://doi.org/10.1103/PhysRevResearch.3.033002} {\bibfield  {journal} {\bibinfo  {journal} {Phys. Rev. Res.}\ }\textbf {\bibinfo {volume} {3}},\ \bibinfo {pages} {033002} (\bibinfo {year} {2021})}\BibitemShut {NoStop}%
\bibitem [{\citenamefont {Foss-Feig}\ \emph {et~al.}(2022)\citenamefont {Foss-Feig}, \citenamefont {Ragole}, \citenamefont {Potter}, \citenamefont {Dreiling}, \citenamefont {Figgatt}, \citenamefont {Gaebler}, \citenamefont {Hall}, \citenamefont {Moses}, \citenamefont {Pino}, \citenamefont {Spaun}, \citenamefont {Neyenhuis},\ and\ \citenamefont {Hayes}}]{FossFeig2022}%
  \BibitemOpen
  \bibfield  {author} {\bibinfo {author} {\bibfnamefont {M.}~\bibnamefont {Foss-Feig}}, \bibinfo {author} {\bibfnamefont {S.}~\bibnamefont {Ragole}}, \bibinfo {author} {\bibfnamefont {A.}~\bibnamefont {Potter}}, \bibinfo {author} {\bibfnamefont {J.}~\bibnamefont {Dreiling}}, \bibinfo {author} {\bibfnamefont {C.}~\bibnamefont {Figgatt}}, \bibinfo {author} {\bibfnamefont {J.}~\bibnamefont {Gaebler}}, \bibinfo {author} {\bibfnamefont {A.}~\bibnamefont {Hall}}, \bibinfo {author} {\bibfnamefont {S.}~\bibnamefont {Moses}}, \bibinfo {author} {\bibfnamefont {J.}~\bibnamefont {Pino}}, \bibinfo {author} {\bibfnamefont {B.}~\bibnamefont {Spaun}}, \bibinfo {author} {\bibfnamefont {B.}~\bibnamefont {Neyenhuis}},\ and\ \bibinfo {author} {\bibfnamefont {D.}~\bibnamefont {Hayes}},\ }\bibfield  {title} {\bibinfo {title} {Entanglement from tensor networks on a trapped-ion quantum computer},\ }\href {https://doi.org/10.1103/PhysRevLett.128.150504} {\bibfield  {journal} {\bibinfo  {journal} {Phys. Rev. Lett.}\ }\textbf {\bibinfo
  {volume} {128}},\ \bibinfo {pages} {150504} (\bibinfo {year} {2022})}\BibitemShut {NoStop}%
\bibitem [{\citenamefont {Anand}\ \emph {et~al.}(2023{\natexlab{b}})\citenamefont {Anand}, \citenamefont {Hauschild}, \citenamefont {Zhang}, \citenamefont {Potter},\ and\ \citenamefont {Zaletel}}]{Anand2022}%
  \BibitemOpen
  \bibfield  {author} {\bibinfo {author} {\bibfnamefont {S.}~\bibnamefont {Anand}}, \bibinfo {author} {\bibfnamefont {J.}~\bibnamefont {Hauschild}}, \bibinfo {author} {\bibfnamefont {Y.}~\bibnamefont {Zhang}}, \bibinfo {author} {\bibfnamefont {A.~C.}\ \bibnamefont {Potter}},\ and\ \bibinfo {author} {\bibfnamefont {M.~P.}\ \bibnamefont {Zaletel}},\ }\bibfield  {title} {\bibinfo {title} {Holographic quantum simulation of entanglement renormalization circuits},\ }\href {https://doi.org/10.1103/PRXQuantum.4.030334} {\bibfield  {journal} {\bibinfo  {journal} {PRX Quantum}\ }\textbf {\bibinfo {volume} {4}},\ \bibinfo {pages} {030334} (\bibinfo {year} {2023}{\natexlab{b}})}\BibitemShut {NoStop}%
\bibitem [{\citenamefont {Lindner}\ and\ \citenamefont {Rudolph}(2009)}]{PhysRevLett.103.113602}%
  \BibitemOpen
  \bibfield  {author} {\bibinfo {author} {\bibfnamefont {N.~H.}\ \bibnamefont {Lindner}}\ and\ \bibinfo {author} {\bibfnamefont {T.}~\bibnamefont {Rudolph}},\ }\bibfield  {title} {\bibinfo {title} {Proposal for pulsed on-demand sources of photonic cluster state strings},\ }\href {https://doi.org/10.1103/PhysRevLett.103.113602} {\bibfield  {journal} {\bibinfo  {journal} {Phys. Rev. Lett.}\ }\textbf {\bibinfo {volume} {103}},\ \bibinfo {pages} {113602} (\bibinfo {year} {2009})}\BibitemShut {NoStop}%
\bibitem [{\citenamefont {Browne}\ and\ \citenamefont {Rudolph}(2005)}]{PhysRevLett.95.010501}%
  \BibitemOpen
  \bibfield  {author} {\bibinfo {author} {\bibfnamefont {D.~E.}\ \bibnamefont {Browne}}\ and\ \bibinfo {author} {\bibfnamefont {T.}~\bibnamefont {Rudolph}},\ }\bibfield  {title} {\bibinfo {title} {Resource-efficient linear optical quantum computation},\ }\href {https://doi.org/10.1103/PhysRevLett.95.010501} {\bibfield  {journal} {\bibinfo  {journal} {Phys. Rev. Lett.}\ }\textbf {\bibinfo {volume} {95}},\ \bibinfo {pages} {010501} (\bibinfo {year} {2005})}\BibitemShut {NoStop}%
\bibitem [{\citenamefont {Economou}\ \emph {et~al.}(2010)\citenamefont {Economou}, \citenamefont {Lindner},\ and\ \citenamefont {Rudolph}}]{PhysRevLett.105.093601}%
  \BibitemOpen
  \bibfield  {author} {\bibinfo {author} {\bibfnamefont {S.~E.}\ \bibnamefont {Economou}}, \bibinfo {author} {\bibfnamefont {N.}~\bibnamefont {Lindner}},\ and\ \bibinfo {author} {\bibfnamefont {T.}~\bibnamefont {Rudolph}},\ }\bibfield  {title} {\bibinfo {title} {Optically generated 2-dimensional photonic cluster state from coupled quantum dots},\ }\href {https://doi.org/10.1103/PhysRevLett.105.093601} {\bibfield  {journal} {\bibinfo  {journal} {Phys. Rev. Lett.}\ }\textbf {\bibinfo {volume} {105}},\ \bibinfo {pages} {093601} (\bibinfo {year} {2010})}\BibitemShut {NoStop}%
\bibitem [{\citenamefont {Pichler}\ \emph {et~al.}(2017)\citenamefont {Pichler}, \citenamefont {Choi}, \citenamefont {Zoller},\ and\ \citenamefont {Lukin}}]{doi:10.1073/pnas.1711003114}%
  \BibitemOpen
  \bibfield  {author} {\bibinfo {author} {\bibfnamefont {H.}~\bibnamefont {Pichler}}, \bibinfo {author} {\bibfnamefont {S.}~\bibnamefont {Choi}}, \bibinfo {author} {\bibfnamefont {P.}~\bibnamefont {Zoller}},\ and\ \bibinfo {author} {\bibfnamefont {M.~D.}\ \bibnamefont {Lukin}},\ }\bibfield  {title} {\bibinfo {title} {Universal photonic quantum computation via time-delayed feedback},\ }\href {https://doi.org/10.1073/pnas.1711003114} {\bibfield  {journal} {\bibinfo  {journal} {Proceedings of the National Academy of Sciences}\ }\textbf {\bibinfo {volume} {114}},\ \bibinfo {pages} {11362} (\bibinfo {year} {2017})}\BibitemShut {NoStop}%
\bibitem [{\citenamefont {Balasubramanian}\ \emph {et~al.}(2023)\citenamefont {Balasubramanian}, \citenamefont {Lake},\ and\ \citenamefont {Choi}}]{balasubramanian20232d}%
  \BibitemOpen
  \bibfield  {author} {\bibinfo {author} {\bibfnamefont {S.}~\bibnamefont {Balasubramanian}}, \bibinfo {author} {\bibfnamefont {E.}~\bibnamefont {Lake}},\ and\ \bibinfo {author} {\bibfnamefont {S.}~\bibnamefont {Choi}},\ }\bibfield  {title} {\bibinfo {title} {2d hamiltonians with exotic bipartite and topological entanglement},\ }\href@noop {} {\bibfield  {journal} {\bibinfo  {journal} {arXiv preprint arXiv:2305.07028}\ } (\bibinfo {year} {2023})}\BibitemShut {NoStop}%
\bibitem [{\citenamefont {Zhang}\ and\ \citenamefont {Klich}(2023)}]{zhang2023coupled}%
  \BibitemOpen
  \bibfield  {author} {\bibinfo {author} {\bibfnamefont {Z.}~\bibnamefont {Zhang}}\ and\ \bibinfo {author} {\bibfnamefont {I.}~\bibnamefont {Klich}},\ }\bibfield  {title} {\bibinfo {title} {{Coupled Fredkin and Motzkin chains from quantum six- and nineteen-vertex models}},\ }\href {https://doi.org/10.21468/SciPostPhys.15.2.044} {\bibfield  {journal} {\bibinfo  {journal} {SciPost Phys.}\ }\textbf {\bibinfo {volume} {15}},\ \bibinfo {pages} {044} (\bibinfo {year} {2023})}\BibitemShut {NoStop}%
\bibitem [{\citenamefont {Zhang}\ and\ \citenamefont {Klich}(2024)}]{Zhang2022quantum}%
  \BibitemOpen
  \bibfield  {author} {\bibinfo {author} {\bibfnamefont {Z.}~\bibnamefont {Zhang}}\ and\ \bibinfo {author} {\bibfnamefont {I.}~\bibnamefont {Klich}},\ }\bibfield  {title} {\bibinfo {title} {{Quantum lozenge tiling and entanglement phase transition}},\ }\href {https://doi.org/10.22331/q-2024-10-10-1497} {\bibfield  {journal} {\bibinfo  {journal} {Quantum}\ }\textbf {\bibinfo {volume} {8}},\ \bibinfo {pages} {1497} (\bibinfo {year} {2024})},\ \Eprint {https://arxiv.org/abs/2210.01098} {arXiv:2210.01098 [quant-ph]} \BibitemShut {NoStop}%
\bibitem [{\citenamefont {Scarpa}\ \emph {et~al.}(2020)\citenamefont {Scarpa}, \citenamefont {Moln\'ar}, \citenamefont {Ge}, \citenamefont {Garc\'{\i}a-Ripoll}, \citenamefont {Schuch}, \citenamefont {P\'erez-Garc\'{\i}a},\ and\ \citenamefont {Iblisdir}}]{PhysRevLett.125.210504}%
  \BibitemOpen
  \bibfield  {author} {\bibinfo {author} {\bibfnamefont {G.}~\bibnamefont {Scarpa}}, \bibinfo {author} {\bibfnamefont {A.}~\bibnamefont {Moln\'ar}}, \bibinfo {author} {\bibfnamefont {Y.}~\bibnamefont {Ge}}, \bibinfo {author} {\bibfnamefont {J.~J.}\ \bibnamefont {Garc\'{\i}a-Ripoll}}, \bibinfo {author} {\bibfnamefont {N.}~\bibnamefont {Schuch}}, \bibinfo {author} {\bibfnamefont {D.}~\bibnamefont {P\'erez-Garc\'{\i}a}},\ and\ \bibinfo {author} {\bibfnamefont {S.}~\bibnamefont {Iblisdir}},\ }\bibfield  {title} {\bibinfo {title} {Projected entangled pair states: Fundamental analytical and numerical limitations},\ }\href {https://doi.org/10.1103/PhysRevLett.125.210504} {\bibfield  {journal} {\bibinfo  {journal} {Phys. Rev. Lett.}\ }\textbf {\bibinfo {volume} {125}},\ \bibinfo {pages} {210504} (\bibinfo {year} {2020})}\BibitemShut {NoStop}%
\bibitem [{Note2()}]{Note2}%
  \BibitemOpen
  \bibinfo {note} {The dimensionality refers to the spatial dimension of the surface and the states.}\BibitemShut {Stop}%
\bibitem [{\citenamefont {Kardar}\ \emph {et~al.}(1986)\citenamefont {Kardar}, \citenamefont {Parisi},\ and\ \citenamefont {Zhang}}]{kardar1986dynamic}%
  \BibitemOpen
  \bibfield  {author} {\bibinfo {author} {\bibfnamefont {M.}~\bibnamefont {Kardar}}, \bibinfo {author} {\bibfnamefont {G.}~\bibnamefont {Parisi}},\ and\ \bibinfo {author} {\bibfnamefont {Y.-C.}\ \bibnamefont {Zhang}},\ }\bibfield  {title} {\bibinfo {title} {Dynamic scaling of growing interfaces},\ }\href {https://doi.org/10.1103/PhysRevLett.56.889} {\bibfield  {journal} {\bibinfo  {journal} {Physical Review Letters}\ }\textbf {\bibinfo {volume} {56}},\ \bibinfo {pages} {889} (\bibinfo {year} {1986})}\BibitemShut {NoStop}%
\bibitem [{\citenamefont {Edwards}\ and\ \citenamefont {Wilkinson}(1982)}]{edwards1982surface}%
  \BibitemOpen
  \bibfield  {author} {\bibinfo {author} {\bibfnamefont {S.~F.}\ \bibnamefont {Edwards}}\ and\ \bibinfo {author} {\bibfnamefont {D.}~\bibnamefont {Wilkinson}},\ }\bibfield  {title} {\bibinfo {title} {The surface statistics of a granular aggregate},\ }\href {https://doi.org/10.1098/rspa.1982.0056} {\bibfield  {journal} {\bibinfo  {journal} {Proceedings of the Royal Society of London. A. Mathematical and Physical Sciences}\ }\textbf {\bibinfo {volume} {381}},\ \bibinfo {pages} {17} (\bibinfo {year} {1982})}\BibitemShut {NoStop}%
\bibitem [{Note3()}]{Note3}%
  \BibitemOpen
  \bibinfo {note} {This means the entropy of subregion $D$ scales faster than the size of $\partial D$ but slower than the size of $D$.}\BibitemShut {Stop}%
\bibitem [{\citenamefont {Morral-Yepes}\ \emph {et~al.}(2024)\citenamefont {Morral-Yepes}, \citenamefont {Smith}, \citenamefont {Sondhi},\ and\ \citenamefont {Pollmann}}]{morral2024entanglement}%
  \BibitemOpen
  \bibfield  {author} {\bibinfo {author} {\bibfnamefont {R.}~\bibnamefont {Morral-Yepes}}, \bibinfo {author} {\bibfnamefont {A.}~\bibnamefont {Smith}}, \bibinfo {author} {\bibfnamefont {S.}~\bibnamefont {Sondhi}},\ and\ \bibinfo {author} {\bibfnamefont {F.}~\bibnamefont {Pollmann}},\ }\bibfield  {title} {\bibinfo {title} {Entanglement transitions in unitary circuit games},\ }\href {https://doi.org/10.1103/PRXQuantum.5.010309} {\bibfield  {journal} {\bibinfo  {journal} {PRX Quantum}\ }\textbf {\bibinfo {volume} {5}},\ \bibinfo {pages} {010309} (\bibinfo {year} {2024})}\BibitemShut {NoStop}%
\bibitem [{\citenamefont {Meakin}\ \emph {et~al.}(1986)\citenamefont {Meakin}, \citenamefont {Ramanlal}, \citenamefont {Sander},\ and\ \citenamefont {Ball}}]{meakin1986ballistic}%
  \BibitemOpen
  \bibfield  {author} {\bibinfo {author} {\bibfnamefont {P.}~\bibnamefont {Meakin}}, \bibinfo {author} {\bibfnamefont {P.}~\bibnamefont {Ramanlal}}, \bibinfo {author} {\bibfnamefont {L.~M.}\ \bibnamefont {Sander}},\ and\ \bibinfo {author} {\bibfnamefont {R.}~\bibnamefont {Ball}},\ }\bibfield  {title} {\bibinfo {title} {Ballistic deposition on surfaces},\ }\href {https://doi.org/10.1103/PhysRevA.34.5091} {\bibfield  {journal} {\bibinfo  {journal} {Physical Review A}\ }\textbf {\bibinfo {volume} {34}},\ \bibinfo {pages} {5091} (\bibinfo {year} {1986})}\BibitemShut {NoStop}%
\bibitem [{\citenamefont {Chame}\ and\ \citenamefont {Reis}(2002)}]{chame2002crossover}%
  \BibitemOpen
  \bibfield  {author} {\bibinfo {author} {\bibfnamefont {A.}~\bibnamefont {Chame}}\ and\ \bibinfo {author} {\bibfnamefont {F.~A.}\ \bibnamefont {Reis}},\ }\bibfield  {title} {\bibinfo {title} {Crossover effects in a discrete deposition model with kardar-parisi-zhang scaling},\ }\href {https://doi.org/10.1103/PhysRevE.66.051104} {\bibfield  {journal} {\bibinfo  {journal} {Physical Review E}\ }\textbf {\bibinfo {volume} {66}},\ \bibinfo {pages} {051104} (\bibinfo {year} {2002})}\BibitemShut {NoStop}%
\bibitem [{\citenamefont {Katzav}\ and\ \citenamefont {Schwartz}(2004)}]{PhysRevE.70.061608}%
  \BibitemOpen
  \bibfield  {author} {\bibinfo {author} {\bibfnamefont {E.}~\bibnamefont {Katzav}}\ and\ \bibinfo {author} {\bibfnamefont {M.}~\bibnamefont {Schwartz}},\ }\bibfield  {title} {\bibinfo {title} {What is the connection between ballistic deposition and the kardar-parisi-zhang equation?},\ }\href {https://doi.org/10.1103/PhysRevE.70.061608} {\bibfield  {journal} {\bibinfo  {journal} {Phys. Rev. E}\ }\textbf {\bibinfo {volume} {70}},\ \bibinfo {pages} {061608} (\bibinfo {year} {2004})}\BibitemShut {NoStop}%
\bibitem [{\citenamefont {Barab{\'a}si}\ and\ \citenamefont {Stanley}(1995)}]{barabasi1995fractal}%
  \BibitemOpen
  \bibfield  {author} {\bibinfo {author} {\bibfnamefont {A.-L.}\ \bibnamefont {Barab{\'a}si}}\ and\ \bibinfo {author} {\bibfnamefont {H.~E.}\ \bibnamefont {Stanley}},\ }\href@noop {} {\emph {\bibinfo {title} {Fractal concepts in surface growth}}}\ (\bibinfo  {publisher} {Cambridge university press},\ \bibinfo {year} {1995})\BibitemShut {NoStop}%
\bibitem [{Note4()}]{Note4}%
  \BibitemOpen
  \bibinfo {note} {More explicitly, $h_i(0)=1$ for $i$ odd, $h_i(0)=0$ for $i$ even, with boundary $h_0(0)=h_{L+1}(0)=0$.}\BibitemShut {Stop}%
\bibitem [{Note5()}]{Note5}%
  \BibitemOpen
  \bibinfo {note} {In \cite {morral2024entanglement}, the height difference constraint and update rules \protect \eqref {eq:dep} and \protect \eqref {eq:eva} are directly given, with the initial condition as a flat surface.}\BibitemShut {Stop}%
\bibitem [{Note6()}]{Note6}%
  \BibitemOpen
  \bibinfo {note} {This is earlier than the equilibration time of KPZ dynamics due to the open boundary condition and the neighboring height difference constraint.}\BibitemShut {Stop}%
\bibitem [{Note7()}]{Note7}%
  \BibitemOpen
  \bibinfo {note} {Around each vertex, the sum of the left two spins should equal the sum of the right two spins.}\BibitemShut {Stop}%
\bibitem [{Note8()}]{Note8}%
  \BibitemOpen
  \bibinfo {note} {{ $p_\alpha $ depends on the entire trajectory $\alpha $, which records the number of sites corresponding to no event, deposition, evaporation, and the special evaporation at $h=0$. As discussed in the rest of the section, under absorbing boundary conditions, where the trajectories are post-selected and reweighted, no explicit closed-form expression exists, while under reflecting boundary conditions $p_\alpha $ is simply the product of the local probabilities specified in the update rules \protect \eqref {eq:UF1}-\protect \eqref {eq:UE6}. For this reason we do not write a single explicit functional form of $p$, the tunable parameter.}}\BibitemShut {Stop}%
\bibitem [{Note9()}]{Note9}%
  \BibitemOpen
  \bibinfo {note} {The two sets of states $\{|w_L\rangle \}$ and $\{|w_R\rangle \}$ are orthonormal bases because of the fixed initial and final conditions, and color matching .}\BibitemShut {Stop}%
\bibitem [{\citenamefont {Yu}\ \emph {et~al.}(1994)\citenamefont {Yu}, \citenamefont {Pang},\ and\ \citenamefont {Halpin-Healy}}]{PhysRevE.50.5111}%
  \BibitemOpen
  \bibfield  {author} {\bibinfo {author} {\bibfnamefont {Y.-K.}\ \bibnamefont {Yu}}, \bibinfo {author} {\bibfnamefont {N.-N.}\ \bibnamefont {Pang}},\ and\ \bibinfo {author} {\bibfnamefont {T.}~\bibnamefont {Halpin-Healy}},\ }\bibfield  {title} {\bibinfo {title} {Concise calculation of the scaling function, exponents, and probability functional of the edwards-wilkinson equation with correlated noise},\ }\href {https://doi.org/10.1103/PhysRevE.50.5111} {\bibfield  {journal} {\bibinfo  {journal} {Phys. Rev. E}\ }\textbf {\bibinfo {volume} {50}},\ \bibinfo {pages} {5111} (\bibinfo {year} {1994})}\BibitemShut {NoStop}%
\bibitem [{Note10()}]{Note10}%
  \BibitemOpen
  \bibinfo {note} {$|\Psi _{\protect \text {ref}}\rangle $ does not have a local parent Hamiltonian because the Hamiltonian has to detect the superposition of configurations with the surface height touching zero with a different update rule. The height is a non-local information, involving an extensive number of spins along a column, and can only be energetically constrained by nonlocal terms in the Hamiltonian.}\BibitemShut {Stop}%
\bibitem [{Note11()}]{Note11}%
  \BibitemOpen
  \bibinfo {note} {We can also extend the procedure to generate $|\Psi _{\protect \text {abs,colored}}\rangle $ by using infinite stacks in both directions. However, the post-selection involves states going below the initial horizon configuration, significantly reducing the efficiency.}\BibitemShut {Stop}%
\bibitem [{Note12()}]{Note12}%
  \BibitemOpen
  \bibinfo {note} {The top row of the state corresponds to the final condition and is not part of the generation either.}\BibitemShut {Stop}%
\bibitem [{\citenamefont {Barbiero}\ \emph {et~al.}(2017)\citenamefont {Barbiero}, \citenamefont {Dell'Anna}, \citenamefont {Trombettoni},\ and\ \citenamefont {Korepin}}]{PhysRevB.96.180404}%
  \BibitemOpen
  \bibfield  {author} {\bibinfo {author} {\bibfnamefont {L.}~\bibnamefont {Barbiero}}, \bibinfo {author} {\bibfnamefont {L.}~\bibnamefont {Dell'Anna}}, \bibinfo {author} {\bibfnamefont {A.}~\bibnamefont {Trombettoni}},\ and\ \bibinfo {author} {\bibfnamefont {V.~E.}\ \bibnamefont {Korepin}},\ }\bibfield  {title} {\bibinfo {title} {Haldane topological orders in motzkin spin chains},\ }\href {https://doi.org/10.1103/PhysRevB.96.180404} {\bibfield  {journal} {\bibinfo  {journal} {Phys. Rev. B}\ }\textbf {\bibinfo {volume} {96}},\ \bibinfo {pages} {180404} (\bibinfo {year} {2017})}\BibitemShut {NoStop}%
\bibitem [{\citenamefont {Wei}\ \emph {et~al.}(2011)\citenamefont {Wei}, \citenamefont {Affleck},\ and\ \citenamefont {Raussendorf}}]{PhysRevLett.106.070501}%
  \BibitemOpen
  \bibfield  {author} {\bibinfo {author} {\bibfnamefont {T.-C.}\ \bibnamefont {Wei}}, \bibinfo {author} {\bibfnamefont {I.}~\bibnamefont {Affleck}},\ and\ \bibinfo {author} {\bibfnamefont {R.}~\bibnamefont {Raussendorf}},\ }\bibfield  {title} {\bibinfo {title} {Affleck-kennedy-lieb-tasaki state on a honeycomb lattice is a universal quantum computational resource},\ }\href {https://doi.org/10.1103/PhysRevLett.106.070501} {\bibfield  {journal} {\bibinfo  {journal} {Phys. Rev. Lett.}\ }\textbf {\bibinfo {volume} {106}},\ \bibinfo {pages} {070501} (\bibinfo {year} {2011})}\BibitemShut {NoStop}%
\bibitem [{\citenamefont {Else}\ \emph {et~al.}(2012)\citenamefont {Else}, \citenamefont {Schwarz}, \citenamefont {Bartlett},\ and\ \citenamefont {Doherty}}]{PhysRevLett.108.240505}%
  \BibitemOpen
  \bibfield  {author} {\bibinfo {author} {\bibfnamefont {D.~V.}\ \bibnamefont {Else}}, \bibinfo {author} {\bibfnamefont {I.}~\bibnamefont {Schwarz}}, \bibinfo {author} {\bibfnamefont {S.~D.}\ \bibnamefont {Bartlett}},\ and\ \bibinfo {author} {\bibfnamefont {A.~C.}\ \bibnamefont {Doherty}},\ }\bibfield  {title} {\bibinfo {title} {Symmetry-protected phases for measurement-based quantum computation},\ }\href {https://doi.org/10.1103/PhysRevLett.108.240505} {\bibfield  {journal} {\bibinfo  {journal} {Phys. Rev. Lett.}\ }\textbf {\bibinfo {volume} {108}},\ \bibinfo {pages} {240505} (\bibinfo {year} {2012})}\BibitemShut {NoStop}%
\bibitem [{\citenamefont {Stephen}\ \emph {et~al.}(2017)\citenamefont {Stephen}, \citenamefont {Wang}, \citenamefont {Prakash}, \citenamefont {Wei},\ and\ \citenamefont {Raussendorf}}]{PhysRevLett.119.010504}%
  \BibitemOpen
  \bibfield  {author} {\bibinfo {author} {\bibfnamefont {D.~T.}\ \bibnamefont {Stephen}}, \bibinfo {author} {\bibfnamefont {D.-S.}\ \bibnamefont {Wang}}, \bibinfo {author} {\bibfnamefont {A.}~\bibnamefont {Prakash}}, \bibinfo {author} {\bibfnamefont {T.-C.}\ \bibnamefont {Wei}},\ and\ \bibinfo {author} {\bibfnamefont {R.}~\bibnamefont {Raussendorf}},\ }\bibfield  {title} {\bibinfo {title} {Computational power of symmetry-protected topological phases},\ }\href {https://doi.org/10.1103/PhysRevLett.119.010504} {\bibfield  {journal} {\bibinfo  {journal} {Phys. Rev. Lett.}\ }\textbf {\bibinfo {volume} {119}},\ \bibinfo {pages} {010504} (\bibinfo {year} {2017})}\BibitemShut {NoStop}%
\bibitem [{Note13()}]{Note13}%
  \BibitemOpen
  \bibinfo {note} {Here we only explicitly write out the part of the stacks that are updated.}\BibitemShut {Stop}%
\end{thebibliography}%
\end{document}